\input harvmac
% macros %%%%%%%%%%%%%%%%%%%%%%%%%%%%%%%%%%%%%%%%%%%%%%%%%%%%%%%%%%%%%
% miscellaneous macros
\def\CA{{\cal A}}

\def\CM{{\cal M}}\def\CO{{\cal O}}\def\CP{{\cal P}}
\def\CS{{\cal S}}\def\CT{{\cal T}}
\def\CV{{\cal V}}

\def\ie{{\sl i.e.}}  \def\eg{{\sl e.g.}}
 \def\etal{{\sl et al.}}

\def\half{{\textstyle{1\over2}}} \def\third{{\textstyle{1\over3}}}

\def\ltsim{\vcenter{\hbox{$<$}\nointerlineskip\hbox{$\sim$}}}
\def\me#1#2#3{{\langle#1\vert#2\vert#3\rangle}}
\def\ME#1#2#3{{\big\langle#1\big\vert#2\big\vert#3\big\rangle}}

\def\gleft{{(1-\gamma_5)}} \def\gright{{(1+\gamma_5)}}

\def\gsqeff{{g^2_{\rm eff}}}
\def\bk{\hbox{$B_K$}}
\def\kbar{{\overline{K^0}}}
\def\sdsdop{{(\bar s\gamma_\mu L d)(\bar s \gamma_\mu L d)}}
% journal macros

\def\NPB#1{{\sl Nucl. Phys.} {\bf B#1}}
\def\NPBPS#1{{\sl Nucl. Phys.} {\bf B} {\sl (Proc. Suppl.)} {\bf #1}}
\def\PRL#1{{\sl Phys. Rev. Lett.} {\bf #1}}
\def\PRD#1{{\sl Phys. Rev. } {\bf D#1}}
\def\PLB#1{{\sl Phys. Lett.} {\bf #1B}}
% lattice conference macros
\def\seillac{
  Int. Symp. {\sl ``Field theory on the Lattice''},
  Proceedings of the International Symposium on Lattice Field Theory,
  Seillac, France, 1987, edited by A. Billoire \etal,
  \NPBPS{4}, (1988)}
\def\fermilab{
  Int. Symp. {\sl ``LATTICE 88''},
  Proceedings of the International Symposium on Lattice Field Theory,
  Fermilab, USA, 1988, edited by A.S. Kronfeld and P.B. Mackenzie,
  \NPBPS{9} (1989)}
\def\capri{
  Int. Symp. {\sl ``LATTICE 89''},
  Proceedings of the International Symposium on Lattice Field Theory,
  Capri, Italy, 1989, edited by N. Cabibbo \etal,
  \NPBPS{17}, (1990)}
\def\talla{
  Int. Symp. {\sl ``LATTICE 90''},
  Proceedings of the International Symposium on Lattice Field Theory,
  Tallahassee, Florida, USA, 1990, edited by U. M. Heller \etal,
  \NPBPS{20}, (1991)}

% redefine \author
\def\author#1#2{
  \bigskip\centerline{#1}
  \smallskip\centerline{\sl #2}}
% redefine \subsec with \sl rather than \it
\def\subsec#1{
  \global\advance\subsecno by1
  \message{(\secsym\the\subsecno. #1)}
  \ifnum\lastpenalty>9000\else\bigbreak\fi
  \noindent{\sl\secsym\the\subsecno. #1}
  \writetoca{\string\quad {\sl\secsym\the\subsecno. #1}}
  \par\nobreak\medskip\nobreak}

% title %%%%%%%%%%%%%%%%%%%%%%%%%%%%%%%%%%%%%%%%%%%%%%%%%%%%%%%%%%%%%%
\Title
{LA UR-91-3522}
{The Kaon $B$-parameter with Wilson Fermions}

% authors %%%%%%%%%%%%%%%%%%%%%%%%%%%%%%%%%%%%%%%%%%%%%%%%%%%%%%%%%%%%
\author
{Rajan Gupta and David Daniel}
{T-8, MS-B285, Los Alamos National Laboratory, Los Alamos, NM 87545}

\author
{Gregory W. Kilcup}
{Physics Department, The Ohio State University, Columbus, OH 43210}

\author
{Apoorva Patel}
{Supercomputer Education and Research Centre
 and Centre for Theoretical Studies}
\centerline{\sl Indian Institute of Science, Bangalore 560012, India}

\author
{Stephen R. Sharpe}
{Physics Department FM-15,
 University of Washington, Seattle, WA 98195}

% abstract %%%%%%%%%%%%%%%%%%%%%%%%%%%%%%%%%%%%%%%%%%%%%%%%%%%%%%%%%%%
\bigskip
We  calculate the  kaon   $B$-parameter  in quenched   lattice  QCD at
$\beta=6.0$ using Wilson fermions at $\kappa=0.154$  and $0.155$.   We
use two kinds of non-local (``smeared'') sources for quark propagators
to calculate the matrix elements between  states of definite momentum.
The use  of smeared sources yields results  with  much  smaller errors
than obtained in   previous calculations with   Wilson  fermions.   By
combining results for $\vec p =(0,0,0)$ and $\vec p =(0,0,1)$, we show
that one can carry out  the non-perturbative subtraction necessary  to
remove the dominant lattice  artifacts induced by  the chiral symmetry
breaking term  in  the Wilson action.  Our  final  results are in good
agreement   with those  obtained  using   staggered fermions.  We also
present results for $B$-parameters of the $\Delta I = 3/2$ part of the
electromagnetic penguin operators, and preliminary results for \bk\ in
the presence of two flavors of dynamical quarks.

% date %%%%%%%%%%%%%%%%%%%%%%%%%%%%%%%%%%%%%%%%%%%%%%%%%%%%%%%%%%%%%%%
% for draft ...
\def\draftdate{
  \number\day\ \ifcase\month
    \or January\or February\or March\or April \or May\or June\or July
    \or August\or September \or October\or November\or December
  \fi \number\year\ (Draft)}
%\Date{\draftdate}
\Date{11 October 1992}

% paper %%%%%%%%%%%%%%%%%%%%%%%%%%%%%%%%%%%%%%%%%%%%%%%%%%%%%%%%%%%%%%

%\baselineskip=1.5\normalbaselineskip % for double spacing

\newsec{Introduction}

Present   calculations  of  weak matrix   elements   in the   quenched
approximation with Wilson  fermions  suffer from  two  main sources of
error: (i) the  signal is poor   and (ii) there  are large
$O(a)$ corrections due to lack of chiral symmetry \ref
\bkroma{
  M.B. Gavela, \etal, \NPB{306} (1988) 677.
} \ref
\bkucla{
  C. Bernard and A. Soni, \capri\ 495.
}.
In this paper we investigate the calculation of the matrix elements of
four-fermion operators between pseudoscalar states, and in particular
\bk.  To improve the signal we calculate the 3-point function by
sandwiching the operator  between kaons produced  by  smeared sources.
This   trick  has been  used  to   obtain  very accurate  results with
staggered fermions~\ref
\bkprl{
  G. Kilcup, S. Sharpe, R. Gupta and A. Patel, \PRL{64} (1990) 25.}.
In order to reduce the $O(a)$ artifacts we use a momentum-subtraction
technique similar to that tried earlier by the ELC collaboration~\ref
\oldtech{G. Martinelli, \capri\ 523.}.
We find  that the combined method reduces  the statistical  errors for
all four-fermion operators we have looked at, and allows us to perform
non-perturbative subtractions for  removing two  of  the three  chiral
symmetry violating terms in \bk.

The  $O(a)$ corrections arise   due  to  mixing between  operators  of
different  tensor  structure induced   by the explicit chiral symmetry
breaking term introduced by  Wilson  to remove  lattice  doublers.  In
principle this   mixing can  be   calculated in  perturbation  theory,
but there are  large  non-perturbative effects at  values  of $g$
used in  lattice calculations.  There are  two approaches to improving
the situation: one   is to work with  an  improved action  so that the
mixing occurs at $O(g^2 a)$ and $O(a^2)$ rather than at $O(a)$~\ref
\bkimp{
  G. Martinelli \etal, Roma preprint no. 850 (1991).
},
and the second is to  devise non-perturbative methods to subtract  off
the lattice artifacts. It is likely that the eventual solution will be
a combination of the two methods.  To this end we demonstrate that the
calculation of matrix  elements    within states of   definite lattice
momentum works for  $\vec p=(0,0,0)$  and  $(0,0,1)$, and  furthermore
that one can reliably  carry out a  non-perturbative subtraction using
these two  values of momentum.   We  use the kaon $B$-parameter as the
testing  ground for two reasons: (a)  there  are very accurate results
available using staggered fermions on the same set of lattices against
which  we may compare our results,  and  (b) there  is  no mixing with
operators of lower dimension.

To make our non-perturbative method work we need two  kinds  of hadron
source: one that  produces hadrons with  zero  momentum and the  other
that couples to all   momenta.   We construct  zero   momentum  hadron
correlators    using  wall  source     quark propagators,   while  the
Wuppertal source~\ref
\guskencapri{
  S. G\"usken \etal, \PLB{227} (1989) 266\semi
  S. G\"usken, \capri\ 365.
}\
propagators   yield hadron correlators   that  have overlap  with  all
momenta.   We have shown in Ref.~\ref
\ushmwilson{
  D. Daniel,  R. Gupta, G.  Kilcup, A.  Patel and  S.  Sharpe,
  \PRD{46} (1992) 3130
}\
that these two kinds of correlators yield reliable signals for both the
amplitude  and the mass  extracted from 2-point correlation functions.
That paper describes in  detail  the lattices  used in the calculation
and details of the quark propagators and hadron  correlators.  It also
contains results for hadron  masses and  decay constants obtained from
2-point  correlation functions.  We use $35$ lattices
of  size $16^3  \times  40$ at $\beta\equiv 6g^{-2}  =6.0$  with quark
propagators calculated at $\kappa=0.154$ and  $0.155$.  The two values
of $\kappa$  correspond to kaons of  mass $M_K=700$~MeV  and $560$~MeV
respectively, using $a^{-1} = 1.9$~GeV.
%%% roughly correspond to $m_q=m_s$ and $0.7m_s$. %%%%%%%%%%%%%%%%%%%%

The most accurate  results for \bk\ at  $\beta=6.0$ have been obtained
with staggered fermions \bkprl:
\eqn\stagresults{
  B_K = \cases{ 0.70 \pm 0.02
                  \; : \; (16^3\times40{\rm\ lattices}) \cr
                  0.70 \pm 0.01
                  \; : \; (24^3\times40{\rm\ lattices}) \cr}.
}\
There are  two  previous  estimates of \bk\   with  Wilson fermions at
$\beta=6.0$.  The results of Bernard and Soni are \bkucla:
\eqn\oldresultsBS{
  B_K = \cases{ 0.83 \pm 0.11 \pm 0.11
                \; : \; (16^3\times40{\rm\ lattices}) \cr
                0.66 \pm 0.08 \pm 0.04
                \; : \; (24^3\times40{\rm\ lattices}) \cr},
}
and that of the ELC collaboration is \bkroma:
\eqn\oldresultsELC{
  B_K = 0.81 \pm 0.16 \pm 0.06
  \; : \; (10^2\times20\times40{\rm\ lattices}) \ ,
}
where the second error is an estimate  of the systematic  error due to
the subtraction of the  bad chiral behavior.   In all calculations the
lattice kaon consisted of two almost degenerate quarks (the ratio $m_s
/   m_u    \leq 3   $).  The   above   results  were   obtained  after
interpolation/extrapolation of the  lattice  data to  a kaon   mass of
$495\   MeV$.  The large spread in   these numbers  and the systematic
errors due    to  bad  chiral behavior  induced  by   the  Wilson term
underscore the need for further improvements and new methods.
%% Recently, the ELC  collaboration has  repeated the calculation
%% using  an improved  Wilson fermion action  at  $\beta=6.0$,  and their
%% result is
%% \bkimp:
%% \eqn\impresultsELC{
%%   B_K = 1.03 \pm 0.05 \pm 0.06
%%   \; : \; (10^2\times20\times40{\rm\ lattices}).
%% }

We also calculate the $B$-parameter for the  $\Delta I = 3/2$ part of
the electromagnetic penguin operators $\CO_7$  and  $\CO_8$.  Previous
calculations with both Wilson~\ref
\uclaLR{
  C. Bernard, T. Draper, G. Hockney and A. Soni, \seillac\ 483.
} \ref
\ELCLR{
  E. Franco, L. Maiani, G. Martinelli and E. Morelli,
  \NPB{317} (1989) 63.
},
and staggered fermions~\ref
\stagLR{
  S. Sharpe, R. Gupta, G. Guralnik, G. Kilcup and A. Patel,
  \PLB{192} (1987) 149.
}\
show  that   reliable results for  the matrix   elements  of these  LR
operators can be obtained in lattice calculations  and that the vacuum
saturation  approximation   (VSA)  provides  a    good  estimate, \ie\
$B^{3/2}_{7,8}  =  1.0  \pm 0.1$.  Our  estimates are   $0.89(4)$  and
$0.93(5)$ respectively, and we find that  the dominant contribution to
the matrix elements of both the LR operators and their VSA  comes from
the pseudoscalar $\otimes$ pseudoscalar ($\CP $) part of the 4-fermion
operator.  Our data show that matrix elements of  $\CP $ are larger by
a factor  of 10  or more  than other tensor   structures  and that the
2-color  loop  contraction  is roughly  three  times  larger than  the
1-color loop.  Furthermore, as the operator $\CP$ is not
suppressed in the  chiral limit, we believe  that VSA  will be  a good
approximation in cases where   the operator or its fierz transform
contains $\CP$ at tree level.

This paper is organized as follows: in Section 2 we review the problem
induced   by the   Wilson  $r$ term   and  our   partial  solution for
subtracting lattice  artifacts.  In Section 3  we describe the lattice
methods and in Section 4 we present our results for  $B_K$.  We make a
comparison  with  earlier   results  obtained   with  both Wilson  and
staggered  fermions in   Section 5.   Section 6  presents  preliminary
results for \bk\ with two flavors of dynamical quarks. The analysis of
the LR operators is given in Section 7  and we end with conclusions in
Section 8.

\newsec{$B_K$ and the problem of bad chiral behavior}

Weak interactions give rise to  mixing between the  $K^0$ and $\kbar$.
The relevant operator in the low energy  effective weak Hamiltonian is
the  $\Delta{S}=2$ four-fermion operator $\sdsdop$,   where we use the
notation $L=\gleft$ and $R=\gright$.  The value  of the matrix element
of this operator between  a $K^0$ and  $\kbar$ at  a  typical hadronic
scale  is severely influenced by strong  interaction  effects.  It has
become customary  to  parameterize this  matrix element by    the kaon
$B$-parameter, \bk, which measures the deviation from its value in the
VSA:
\eqn\BKdef{
  \ME{\kbar}{\sdsdop}{K^0} = {8\over3} f_K^2 M_K^2 B_K,
}
where $()$ indicates  a trace  over the spin  and color  indices.  The
normalization   used   for   the  decay   constant    is   such   that
$f_\pi=132$~MeV.   If the VSA  is   exact then  $B_K=1$.  To calculate
$B_K$ from first  principles we must turn to  non-perturbative methods
such as  the  lattice.  Our lattice  calculation of \bk\ uses Wilson's
formulation for fermions.   The inherent violation of  chiral symmetry
in this approach leads to technical difficulties which we now review.

To begin with, note that $\sdsdop$ is a special case of the operator
\eqn\Oplusdef{
  \CO_+ = {1\over2}
  \big( (\bar\psi_1 \gamma_\mu L \psi_2)
        (\bar\psi_3 \gamma_\mu L \psi_4)
        + (2 \leftrightarrow 4) \big),
}
with  $\psi_1=\psi_3=s$  and  $\psi_2=\psi_4=d$.   The significance of
this   is that  with  a  chirally  invariant  regulator  $\CO_+$    is
multiplicatively renormalized.  With Wilson fermions, however, this is
not the case: there is mixing of this $LL$ operator  with other tensor
structures in addition    to  an  overall renormalization,    and this
complicates  the  definition of  a lattice operator  with  the desired
continuum behavior. In perturbation theory, the corrected operator has
been calculated to 1-loop in Refs.~\ref
\pertmarti{
  G. Martinelli, \PLB{141} (1984) 395.
}\
and \ref
\pertucla{
  C. Bernard, T. Draper and A. Soni, \PRD{36} (1987) 3224.
}:
\eqn\pertop{
  \CO_+^{\rm cont} = \big(1 + {g^2 \over {16\pi^2}} Z_+(r,a\mu)  \big)
                     \CO_+^{\rm latt} +
                     {4g^2 \over {16\pi^2}} \, r^2 Z^*(r)
                     \big( \CO_+^{STP} + \CO_+^{VA} + \CO_+^{SP} \big)
}
where
\eqn\mixopdef{\eqalign{
  \CO_+^{STP} &= {N-1 \over 16N} \big[
      (\CS + \CT + \CP) + (2\leftrightarrow 4) \big], \cr
  \CO_+^{VA}  &= - {N^2 + N-1 \over 32N} \big[
      (\CV - \CA) + (2\leftrightarrow 4) \big], \cr
  \CO_+^{SP}  &= - {1 \over 16N} \big[
      (\CS - \CP) + (2\leftrightarrow 4) \big], \cr
}}
and $N=3$ is the number of colors.  We have  used a condensed notation
for the allowed Lorentz tensor structures:
\eqn\lstruct{\eqalign{
  \CS &=               (\bar\psi_1                  \psi_2)
                       (\bar\psi_3                  \psi_4), \cr
  \CV &=               (\bar\psi_1\gamma_\mu        \psi_2)
                       (\bar\psi_3\gamma_\mu        \psi_4). \cr
  \CT &= \sum_{\mu<\nu}(\bar\psi_1\sigma_{\mu\nu}   \psi_2)
                       (\bar\psi_3\sigma_{\mu\nu}   \psi_4), \cr
  \CA &=               (\bar\psi_1\gamma_\mu\gamma_5\psi_2)
                       (\bar\psi_3\gamma_\mu\gamma_5\psi_4), \cr
  \CP &=               (\bar\psi_1\gamma_5          \psi_2)
                       (\bar\psi_3\gamma_5          \psi_4), \cr
}}
where $\gamma_\mu,\gamma_5$  are   hermitian  and $\sigma_{\mu\nu}   =
(\gamma_\mu\gamma_\nu-\gamma_\nu\gamma_\mu)/2$.    We  note  that  the
Fierz  transform eigenstates  appearing  in   Eq.  \pertop\ are   only
$(\CV+\CA)$,  $\half(\CV-\CA)\pm(\CS-\CP)$ and $(\CS+\CT+\CP)$;  there
is also no mixing between the fifth eigenstate  of the Fierz transformation
$(\CS-\third\CT+\CP)$ and the operator $\CO_+$ at 1-loop.  There is no
mixing  with lower dimensional  operators, for the  simple reason that
there are {\sl no}  $\Delta{S}=2$  operators of lower  dimension.   We
shall   henceforth   denote  the  perturbatively  corrected  $\sdsdop$
operator (cf.  Eq. \pertop) by $\hat\CO$.

The renormalization coefficients for Wilson  parameter $r=1$ are given
in  Table 1 in  three schemes: the dimensional reduction ($DRED$) used
by Altarelli \etal\
\ref\altarelli{G. Altarelli, G. Curci, G. Martinelli and S. Petrarca,
\NPB{187} (1981) 461}\ and Martinelli \pertmarti,
as well as the ``naive'' dimensional
regularization ($NDR$)  and  the  dimensional reduction ($DR(\overline
{EZ})$)   scheme   used by  Bernard \etal\  in  \pertucla.  A detailed
description of $DRED$ and $NDR$ schemes  and their relative advantages
and disadvantages is given in Ref.~
\ref\burasweisz{A. Buras and P. Weisz, \NPB{333} (1990) 66}.
We tabulate the  relevant results in order to  provide easy reference,
and to allow the reader  to make a rough  estimate of the magnitude of
the scheme dependence. All our results are given in the $DRED$ scheme,
except when we compare raw lattice numbers against those in Ref.~\ref
\bsprivate{
  C. Bernard and A. Soni, private communications.
},
in which case we use $DR(\overline {EZ})$.

For each  of the four-fermion  operators, $\CS$, $\CV$,  $\CT$, $\CA$,
and $\CP$, there  are  two  distinct contractions with  the   external
states.  In the first each bilinear is  contracted with an incoming or
outgoing  kaon corresponding  to two  spin and two  color  traces.  We
label these  contractions by $\CP^2$, $\CS^2$,   $\CV^2$,  $\CA^2$ and
$\CT^2$.  The other contraction consists of  a single  spin  and color
trace which we Fierz transform to two spinor loops.  We
label them  by $\CP^1$,  $\CS^1$, $\CV^1$, $\CA^1$  and $\CT^1$, since
they have a  single  color trace.  We  will find it useful to  further
split the $\CV$, $\CA$  and $\CT$  terms  into their  space  and  time
components,  and denote  these  components by  subscripts $s$ and  $t$
respectively.  This  notation is  similar to  that used with staggered
fermions \bkprl\ and will facilitate later  comparison  of results for
individual operators between the two formulations.

In order to extract \bk, we calculate,  at non-zero momentum transfer,
the matrix element
\eqn\MKdef{
 \CM_K(\vec p) = \ME{\kbar(\vec p)}{\hat\CO(\vec p)}{K^0(\vec p=0)}.
}
In  chiral  perturbation  theory  $\CM_K(\vec  p)$  behaves  as $ \sim
\gamma_K  p_K\cdot  p_{\bar K}$, where  $\gamma_K=8/3\,f_K^2 B_K$, and
$p_K$ and $p_{\bar K}$ are the  on-shell  four-momenta of the external
states, so that  $p_K\cdot p_{\bar  K}=M_K\sqrt{M_K^2 + (\vec  p)^2}$.
Unfortunately, on the lattice  with Wilson fermions chiral symmetry is
explicitly broken and the expansion becomes
\eqn\chiralbehavior{
  \CM_K(\vec p) =
  \alpha + \beta M_K^2 + (\gamma + \gamma_K) p_K \cdot p_{\overline K}
  + \dots .
}
Here  the terms  proportional  to $\alpha$, $\beta$   and $\gamma$ are
unphysical  contributions  arising  due to the  $r$-term in the Wilson
action, and  suppressed  by  one  power of  the  lattice spacing  $a$.
Similar formulae  hold for each   individual spin-color term described
above, and apply to both the on-shell ($\me{\kbar}{\hat\CO}{K^0}$) and
the off-shell ($\me{0}{\hat\CO}{K^0K^0}$) matrix elements.

Using $\hat\CO$ should reduce  the lattice artifacts, but it  will not
eliminate them completely because it  is only  an approximation to the
operator  with the  desired  continuum  behavior.   In particular, the
coefficients $Z$ contain terms of $\CO(g^4)$  and higher that have not
been calculated, and,  more importantly, as previous calculations have
shown, there are large non-perturbative effects.  We therefore require
non-perturbative  methods   to   isolate   the   physical  coefficient
$\gamma_K$.

The most troublesome of the lattice artifacts is $\alpha$.  Failure to
correctly subtract this contamination will mean that \bk\ will diverge
in the chiral limit. To eliminate this, it is not necessary to work at
non-zero   momentum transfer.   One  can simply calculate  $\CM_K(\vec
p=0)$  at different values  of $\kappa$ (that  is, different values of
$M_K$) and take a difference, leaving
\eqn\kappasub{
  \CM_K(\kappa,\vec{p}=0) - \CM_K(\tilde\kappa,\vec{p}=0)
  = (\beta+\gamma+\gamma_K) (M_K^2 - \tilde M_K^2).
}
To  remove $\beta$  one takes  the  difference   of the on-shell   and
off-shell matrix elements.  This method has been used in Refs.~\bkroma\
and  \bkucla,  and suffers from the lack  of  control over final state
interactions between the kaons in the  off-shell amplitude.   A review
of the status of previous results is given in Refs.~\ref
\claudefermilab{
  C. Bernard and A. Soni, \fermilab\ 155.
}\
and \ref
\sharpecapri{
  S. Sharpe, \capri\ 146.
}.

We advocate using momentum  subtraction which eliminates both $\alpha$
{\sl and} $\beta$   at a fixed value of   $M_K$, using  only  on-shell
quantities.   For  example, by  calculating  the  matrix   element  of
$\hat\CO$  for  two different  values  of  $\vec  p$  and  taking  the
difference one gets
\eqn\mediff{
  \CM_K(\vec p) - \CM_K(0) =
    (\gamma + \gamma_K) M_K (E(p) - M_K) + \dots.
}
In practice we calculate
\eqn\bkdiff{
  B_K = {{E(p)\,B_K^L(p) - M_K\,B_K^L(0)} \over {E(p) - M_K}}
        = (\gamma + \gamma_K) + \dots
}
at each  value of $\kappa$, where  by $B_K^L(p)$  we mean the ratio of
the matrix element to its VSA value, both calculated on the lattice at
appropriate momentum transfer.

This  momentum  subtraction scheme  does   not eliminate   the lattice
artifact $\gamma$. Furthermore, in working at  non-zero momentum there
is a danger that higher order terms (for  example quartic  in momenta)
omitted in Eq.~\chiralbehavior\ may  become significant.  Therefore in
order to compare the  lattice result  with  experiment we have to make
the  following  assumptions:  (i)  using  the  perturbatively improved
operator $\hat\CO$ makes $\gamma$ negligible; (ii)  the terms of order
$p^4$ and higher that we have neglected in the chiral expansion do not
have large coefficients.  The first  assumption is  expected to become
more reliable  on using  an improved  Wilson action,  while the second
will come  under better control as  calculations  are   done on larger
spatial lattices  since then   the gap between lattice    momenta will
decrease.  At this stage  the only justification for these assumptions
is the  a posteriori agreement  of results  with those  obtained using
staggered fermions.

\newsec{Methodology}

Our  method   for   calculating  \bk\  requires    that we double  the
$16^3\times40$ lattices in the time  direction, so   that they are  of
size $16^3\times80$.  On  these doubled lattices we   construct hadron
correlators  such that  the  correlator on  time  slices 1--39  is the
forward moving particle  with the source  at  time slice 0, while  the
correlator on time slices 79--41 is the backward  moving particle with
the periodically reflected  source  on  time slice 80.    To calculate
matrix elements we insert  the operator between these  ``forward'' and
``backward''  moving  particles  on  the  original  $16^3   \times 40$
lattices.

In practice, we divide the correlators for the various matrix elements
by  the  product of kaon correlators,  so that we directly  obtain the
$B$-parameters for the various operators $\CO$
\eqn\bkratiodefinition{
  B_\CO \equiv {3 \over 8} \,
  {\me{\kbar(\vec p)}{\CO}{K^0(\vec p =0)} \over
   \me{\kbar(\vec p)}{A_4}{0}\me{0}{A_4}{K^0(\vec p =0)} }.
}
We  can   select the  kaon  momenta by our choice  of   source  and by
inserting momentum into the  operator $\CO$.   The  statistical errors
are reduced because we can average the operator  location over  a time
slice of the lattice.  Away from the  sources, only the lightest state
contributes to the correlators,  and we should find a time-independent
plateau  giving $B_\CO$.  This  method is similar  to the one  we have
used successfully with staggered fermions \bkprl.

The physical picture of  the process  for  calculating matrix elements
using smeared sources is as follows:  a wall  source at $t=0$ produces
zero momentum  $K^0$  which propagates for a time  $t$, at which point
the operator inserts momentum $\vec p$, and the resulting $\kbar$ with
momentum   $\vec  p$ then  propagates  the  remaining $(N_t$-$t)$ time
slices until it is destroyed by a Wuppertal source.  Three factors are
essential for our method to work:
\item{(i)}
The wall  source creates only zero-momentum  kaons; otherwise there is
contamination from matrix elements of kaons with other momenta.
\item{(ii)}
The Wuppertal source has  significant  overlap  with  the   lowest few
momenta allowed on the lattice.
\item{(iii)}
For matrix elements involving $\vec p\ne0$ kaons, we must  ensure that
there exists an overlap region  for the kaons  where a plateau can  be
observed in the $B$-parameter signal.   Thus it is essential that  the
signal for the zero-momentum kaon produced by  the wall source extends
across  the lattice to  the region  where there  is  a  signal for the
non-zero momentum kaon produced by the Wuppertal source.
\par\noindent
In Ref.~\ushmwilson\ we showed  that these conditions are satisfied by
the Wuppertal and wall correlators,  when we use $\vec{p}=(0,0,0)$ and
$\vec{p}=(0,0,1)$.  Furthermore,  there are   a number of  consistency
checks we make:
\item{(1)}
The $\vec p=0$ matrix element  is calculated three different ways; (a)
using wall  sources on both  sides, (b) using wall  source on one side
and Wuppertal source on the other,  and (c) using Wuppertal sources on
both sides (in this case there is a small contamination from the $\vec
p=(0,0,1)$ state).
\item{(2)}
We use two kaon source operators: $\gamma_5$ and  $A_4$.   The plateau
in each  individual $B_\CO$  is  reached  from opposite directions for
these two.  The two results should converge to the same value.
\par\noindent
As  shown  in Tables 2a--3b,  these  checks are satisfied  by our data
within the statistical accuracy.  We  also find that the $B$-parameter
for the operators $\CA_t^2$ and $\CP^2$ are within a factor  of two of
their VSA  values.  As in  the case of staggered fermions,  the  final
value of \bk\ is obtained after a large cancellation between the $\CA$
and $\CV$ components showing that VSA is not a good approximation.

\newsec{Results for $B_K$}

Our final lattice result at a given value of $\kappa$  and $\vec p$ is
obtained  from the  perturbatively  improved  combination (using   the
convention  that all  four quarks have  distinct flavor labels so that
each term has just one Wick contraction)
\eqn\finalform{\eqalign{
  Z_A^2 B_K^L(\vec p)
  = &{ \bigg(1 + {g^2 \over {16\pi^2}} Z_+(r,a\mu) \bigg) }
           \bigg[ \CV^1  +  \CV^2 + \CA^1  +  \CA^2 \bigg] \cr
           + &  {{ g^2 \over {16\pi^2}}\,{{r^2 Z^*(r) } \over 12}}
                      \bigg[ (26\CS^1 +  2 \CS^2)
                           - (18\CP^1 -  6 \CP^2) \cr
                  & \qquad\qquad\qquad
                         + 4 (  \CT^1 +    \CT^2)
                         +   (  \CV^1 -    \CA^1)
                         - 11(  \CV^2 -    \CA^2)    \bigg]. \cr
}}
For simplicity, we have here  used the  operator symbol to  denote its
$B$-parameter.     The   1-loop     perturbative   results   for   the
renormalization constants $Z_A$ and $Z_+$ are given in Table 1.   Note
that  the finite part  of  the  renormalization factor  $(1+{g^2 \over
16\pi^2} Z_+)$ is largely canceled by $Z_A^2$ for $g^2\ltsim1$.

We give  the results for  the  individual $B_\CO$ (without   any $g^2$
corrections) at $\kappa=0.154$ and  $0.155$ in Tables 2a--3b.  In  all
cases we   find that  the   signal  in the  ratio  of   correlators is
significantly better with the operator $\gamma_5$ as  the  kaon source
than with the operator $A_4$, even though the two sets give consistent
results. As  an  example,  we show a comparison  of the two signals in
Figs.   1a  and  1b.  Our final results  therefore  use the $\gamma_5$
numbers.  We point out that in case of non-zero momentum transfer, the
signal for $B_\CO$ only exists closer to the Wuppertal source (at time
slice ``40'')  than  the wall source  (at  time-slice ``0'').  This is
because the  signal in  the  $\vec{p}=(0,0,1)$  kaon  correlators only
extends  for about 20 time-slices.   The overall quality of the signal
for $B_K^L$ (with $g^2=1, \mu a =1.0$)  is shown in  Figs. 1, 2, 3
and 4 at various values of $\kappa$ and $\vec{p}$.

Tables    2a--3b show  that   all   individual   $B_\CO$,   except for
$B_{\CA_t}$, increase by a  factor  of about 2  between $\kappa=0.154$
and $0.155$.  This increase is largely due to the change in  VSA, \ie\
the factor $f_K^2 M_K^2$ decreases by  approximately $1.9$ between the
two $\kappa$ values \ushmwilson.  This shows  that  at these values of
$\kappa$,  the lattice matrix  elements are dominated  by the constant
term $\alpha$.

The contribution of the mixing terms to  $B_K^L$ can be  large only if
the matrix   elements  are large,    since   the  perturbative  mixing
coefficient  is  $\approx 0.005$ for  $g^2=1$.  The data show that the
largest matrix elements are  of the operator  $\CP$; however their net
contribution to  $B_K^L$  is very small, since   $\CP^2  \sim 3 \CP^1$
(approximate VSA).  Both $\CT^2$  and $\CV^2$  are close to zero.  The
next  largest  contribution comes  from  $4\CT^1$, which  is partially
canceled by $26\CS^1+2\CS^2$.  The net result of these features in the
data is  that the contribution of  mixing terms to \bk\  is in fact of
the order of a few percent.   Unfortunately, since the unphysical term
$\gamma$  in  Eq.~\chiralbehavior\  also gets   contributions from the
diagonal operators,  the small  value of  the mixing   terms does  not
provide a bound on $\gamma$.

Given $B_K^L(\vec   p)$  we   calculate \bk\   and  the  errors  using
Eq.\bkdiff\ two ways:  (1) for  each jackknife sample we first perform
the momentum subtraction  and then  the mean value and  the error  are
obtained as  the jackknife estimate over the   35 samples, and (2)  we
construct  the four  quantities  needed  in Eq.~\bkdiff\ independently
along with their errors, and obtain the  final error estimate assuming
that the individual  estimates  are uncorrelated.   Our quoted results
use  the  first  method,  but we  note  that both  the  methods  yield
consistent estimates.

We  have  calculated   $B_K^L(\vec{p}=(0,0,0))$ three different  ways:
using Wuppertal-Wuppertal     ($S-S$),  Wuppertal-wall  ($S-W$),   and
wall-wall ($W-W$) correlators.   For   example,  our results,    using
$g^2 = 1.0$ and $\mu a =1.0$, are:
\eqn\compsrc{\eqalign{
  B_K^L(\kappa=0.154) &=
  \cases{ 0.38(7) \; : \; S-S \cr
          0.37(4) \; : \; S-W \cr
          0.37(4) \; : \; W-W \cr}, \cr
  B_K^L(\kappa=0.155) &=
  \cases{ 0.10(11)  \; : \; S-S \cr
          0.11 (7)  \; : \; S-W \cr
          0.13 (7)  \; : \; W-W \cr}. \cr
}}
The  consistency of  the data suggests that the  contamination  in the
$S-S$ result  from higher  momentum  kaon states  is   at  most  a few
percent.   Since  only the $S-W$ correlators   give  results   at both
$\vec{p}=(0,0,0)$ and $(0,0,1)$, we shall henceforth quote results for
$B_K^L(\vec{p})$ obtained from these.

In order to extract the continuum result for $B_K$ we must
choose the values of both $g^2$ and $\mu a$ to use in Eq.~\finalform.
Lepage and Mackenzie have suggested~\ref
\effgsq{
  G.P. Lepage and P.B. Mackenzie, \talla\  173;
   Fermilab preprint FERMILAB-PUB-19/355-T (9/92)
}\
that perturbation  theory   is much better behaved   if one   uses the
coupling constant in a continuum scheme such as $\overline{MS}$, instead of
the bare lattice $g^2$. They also give a prescription for choosing the
appropriate scale of the   coupling constant.  In  general  this scale
will differ for  the   various operators that   mix  with   $\CO_+$ in
Eq.~\finalform.  To simplify the  calculation we  take all the scales,
and thus all  the  coupling constants, to  be the same, $i.e.$  all of
$O(\pi/a)$.   Then the   Lepage-Mackenzie  prescription amounts to   a
replacement   of the  bare lattice  $g^2$  with an effective  coupling
$\gsqeff\approx1.75 g^2$  at $\beta=6.0$.  To study the  dependence on
$\gsqeff$, we use four different values, $\gsqeff=0.0$, $1.0$, $1.338$
and $1.75$.  It is important to realize that only a 2-loop calculation
of the perturbative  coefficients can test  whether a given choice for
$\gsqeff$ is reasonable. Such calculations that have been done to date
support Lepage-Mackenzie prescription for choosing $\gsqeff$ \effgsq.

The choice of $\mu a$  is of a  different character to  that of $g^2$.
In physical  matrix elements  (e.g.  that  related  to CP-violation in
$K^0-\bar {  K^0}$  mixing) $B_K$  always appears  multiplied    by  a
coefficient function,   such that the  combination  is  independent of
$\mu$.  At leading order, the scale independent combination is
\eqn\anomolousdimension{
\widehat{B}_K = \alpha_s(\mu)^{-6/(33-2N_f)} B_K(\mu) \ ,
}
where $N_f$ is the number of active flavors.  In fact, $\widehat{B}_K$
does have some dependence on $\mu$, coming from the following sources.
First, since we are using only the leading order expression for $Z(\mu
a)$,  $\widehat{B}_K$ does  depend  on  $\mu$ at   non-leading  order:
$d\ln\widehat{B}_K/d\ln\mu \propto g^4(\mu)$.  This is  likely to be a
small effect, and it can  probably be pushed to next  order  given the
fact  that  the  two-loop  anomalous dimension  and  one-loop matching
coefficients are known.  We say "probably" because it is possible that
there  are some residual subtleties  with Wilson  fermions associated
with  the mixing of  $\CO_+$  with  opposite chirality  operators.   A
related source of $\mu$ dependence occurs when $\mu a$ differs greatly
from  unity: then higher order  terms,  proportional  to $[g^2 \ln(\mu
a)]^n$, which are not included  in Eq.~\finalform, become large.  What
is happening is that the  leading  logarithms, which have  been summed
into the coefficient  function, are  partially  incorporated into  the
perturbative  coefficients.  Once again, one  can probably  take these
into account knowing the anomalous  dimension  to  2-loops, or finesse
the problem by  taking $\mu a\sim1$.   Finally, we are calculating the
lattice result in the quenched approximation, for which the  number of
active flavors is zero, while we wish to match to the full theory with
$N_f$ active flavors. This introduces a small $\mu$ dependence.

Our emphasis in  this paper is on  improving methods  for  calculating
$B_K$, and not on extracting final numbers  for $\widehat{B}_K$.  Thus
we choose to quote our results  for a variety  of values of $\mu a$ so
as to allow others some flexibility if  they wish to  use our numbers.
We use  $\mu  a =  1.0, \ \pi,  $  and $1.7 \pi$.    We have a  slight
preference for  $\mu  a=\pi$, since  then  the  continuum and  lattice
cut-offs are matched.

Table 4 shows  the   sensitivity of  the  results  to the  choices  of
parameters, for both $B_K^L$ and the momentum subtracted $B_K$.  There
is a significant variation of the results  with $\gsqeff$ and $\mu a$.
For   a fixed  value  of  $\mu  a$,  the   lattice    results for both
$B_K^L(\vec{p}=(0,0,0))$  and $B_K^L(\vec{p}=(0,0,1))$ increase  as  a
function of $\gsqeff$ due to the increased contribution  of the mixing
operators.  For fixed $\gsqeff$, an increase  in $\mu a$ decreases the
contribution    of   the        diagonal   operators    (note     that
$B_K^L(\vec{p}=(0,0,0))$ at $\kappa = 0.155$ is insensitive to changes
in $\mu a$ because the diagonal contribution happens to be almost zero
there).

As for  $B_K$ after  momentum subtraction,  the estimate  decreases by
$10-20\%$, at both  values of quark  mass, between  $\gsqeff =1.0$ and
$\gsqeff=1.75$, for fixed $\mu a $.  It turns  out that almost all the
variation  comes   from  the   diagonal renormalization constants (for
example, the ratio of $(1+(g^2/16\pi^2)Z_+)$  to $Z_A^2$ changes  from
$0.93$ to $0.8$), and not from operator mixing.
%%
%% For ratio of  $1+(g^2/16\pi^2)Z_+$ to $Z_A^2$ etc see zfac.f
%%
As our present best estimates for \bk\ we quote the results at
$\gsqeff = 1.75 $, and use $\mu a = \pi$:
\eqn\finalnumbers{\eqalign{
  B_K(\kappa=0.154) =  0.68(22) \ ,  \cr
  B_K(\kappa=0.155) =  0.57(23) \ .  \cr
}}

\newsec{Comparison with previous calculations}

The staggered fermion results for \bk\ \bkprl\  are  statistically the
most  accurate and have the correct  chiral behavior.  At $\beta=6.0$,
the kaon mass roughly  matches   for  staggered $m_q=(0.02+0.03)$  and
$\kappa=0.154$, and for staggered $m_q=(0.01+0.02)$ and $\kappa=0.155$
\ref\staghm{
  R. Gupta, G. Guralnik, G.W. Kilcup, S. R. Sharpe,
  \PRD{43} (1991) 2003.
}.
Thus   we  can  compare  the corresponding data for  \bk.    This is a
particularly  good  comparison because the two  calculations have been
done using the  same set of  background gauge  configurations.   Using
$g^2   = 0$ in the  4-fermion renormalization constants, the staggered
results are $0.76(1)$ and  $0.72(2)$ to be  compared with our  numbers
$0.83(21)$  and $0.77(19)$ respectively.  A  striking feature is  that
the errors with Wilson  fermions are a  factor  of 10 or  more larger.
The data in Table 4 indicate that  the errors in individual $B_K^L(p)$
are larger by  about a factor of four,  and the remaining factor comes
from the momentum subtraction.  The staggered results were obtained by
making fits without including the full covariance matrix, and if we do
the same for Wilson fermions then  the errors in individual $B_K^L(p)$
are  reduced by a  factor of about  two.  Thus, at   the level of  the
signal in the correlators, the smeared sources work almost as well for
Wilson fermions   as for   staggered fermions.  It   is the process of
momentum   subtraction that leads  to   a significant increase  in the
error.

It could be argued that  one should actually compare staggered numbers
with our  preferred results  using  $\gsqeff=1.75$ and $\mu  a = \pi$,
\ie\ $0.68(22)$ and $0.57(23)$ respectively. The rationale for this is
that a  good  choice of  $\gsqeff$ and $\mu   a $  aims to reduce  the
effects  of artifacts $\alpha,\  \beta$  and $\gamma$.   Recall   that
$\alpha$   and  $\beta$ are  absent  for staggered   fermions  while a
$\gamma$ like  term exists  due  to  operator normalization. With this
choice  the Wilson  fermion  estimates  lie  systematically below  the
staggered  values.    The  perturbative  corrections   for   staggered
fermions,  though  small, have  not  been  included  in the  published
results of $B_K$~\ref
\srsbook{
  S. Sharpe, in {\it Standard Model, Hadron Phenomenology
  and Weak Decays on the Lattice},
  Ed. G. Martinelli, World Scientific.
}.
Including  them would  decrease \bk\   by $\approx(2\gsqeff )\%$.  The
remaining  difference could be  due to the difference  in   the $O(a)$
corrections for  the   two fermion   formulations, \eg\  the  artifact
$\gamma$.  In any  case,  it is not clear  whether  the  difference is
significant, given the size  of the errors in  the Wilson results  and
the large variation with $\gsqeff$ and $\mu a$.

We can also make a direct comparison with  results obtained by Bernard
and  Soni  using  Wilson   fermions  \bkucla.   They   have calculated
$B_K^L(\vec{p}=0)$  using  the perturbatively improved  operator  on a
subset   of the same   lattices (they used  only  19  lattices as they
skipped every other one in each of the  two  streams), and at the same
two  values  of $\kappa$.   Their method  of  extraction of $B_K^L$ is
described in  Ref.~\bkucla,  and is different  than the method we have
used.  Using $\gsqeff = 1.338 $ and  $\mu  a = 1.7 \pi$  their results
are  $B_K^L(\vec{p}=0)=0.36(22)$ and  $0.24(45)$ at $\kappa=0.154$ and
$0.155$ respectively~\bsprivate,
to be compared with  $B_K^L(\vec{p}=0)=0.38(5)$ and $0.11(7)$ obtained
by us.  Even after allowing   for a factor   of $\sim\sqrt 2$  due  to
statistics, it is clear on the basis  of  this comparison that our use
of smeared sources has reduced the errors considerably.

Though our results  for $B_K^L$ are  more  accurate compared to  those
obtained by other groups using Wilson fermions,  our final results for
\bk\ have larger errors (cf. Eq. \oldresultsBS,  \oldresultsELC).  The
gain due to the use of smeared sources is  compensated by the increase
in error due to momentum subtraction.  The advantage of using momentum
subtraction  is   that it  unambiguously  removes   lattice  artifacts
$\alpha$ and $\beta$. Also, numerical errors in $B_K^L(\vec{p}=1)$, as
well as  the contribution of  quartic terms  in  the chiral expansion,
should decrease when using a larger lattice due to the decrease in the
value of lattice momenta.

One  further  qualitative comparison  that   we  can make is  for  the
$B$-parameters  (without perturbative  improvements but after momentum
subtraction) of the individual space/time and 1-loop/2-loop components
of the four-fermion vector and axial operators, with the corresponding
results obtained using  staggered fermions \bkprl.  Such  a comparison
is  possible  because, as discussed above,   the  effects  of operator
mixing  are  small.   This   comparison  provides  information  on the
reliability of the momentum subtraction procedure for Wilson fermions.
Furthermore, as explained  in  Ref.~\ref
\srsclogs{S. Sharpe, \PRD{46} (1992) 3146},
the chiral behavior of $B_\CV$ and $B_\CA$ is known; both are expected
to  increase in  magnitude with decreasing  quark mass due  to  chiral
logarithms and finite volume  dependence, and can therefore provide  a
sensitive test at small quark masses.   The  results of our comparison
are shown in Table 5.  Though  the  errors in the results  with Wilson
fermions are  much larger, it is   reassuring to see that  the central
values  are in good agreement.   In  fact the   agreement is far  more
impressive than the errors would naively lead us to  believe.  We need
to   perform calculations  at more values   of $\kappa$ and $\beta$ to
confirm this favorable behavior.

\newsec{Results with two flavors of dynamical fermions}

We have  also estimated \bk,  using the  same methodology as above, on
$16^4$ lattice configurations  generated with two flavors of dynamical
Wilson quarks.  The details of these lattices are given in Ref.~\ref
\wftwo{
  R. Gupta, C. Baillie, R. Brickner,  G. Kilcup,
  A. Patel and S. R. Sharpe, \PRD{44} (1991)  3272.
}.
The kaon  mass at  $\beta=5.5$, $\kappa=0.159$  and $0.160$ is roughly
$M_K=860$~MeV   and    $650$~MeV  respectively;  and  at  $\beta=5.6$,
$\kappa=0.156$  and  $0.157$ it   is  about $1050$~MeV  and  $820$~MeV
respectively.
%% $\beta=5.5$ roughly correspond to $m_q=1.5m_s$ and $0.9m_s$;
%% $\beta=5.6$ roughly correspond to $m_q=2.3m_s$ and $1.4m_s$.
This calculation of \bk\ has been done only with  kaons created with a
$\gamma_5$ operator.  We  find that a  time interval of  $16$ is large
enough to give a  stable plateau over  about 6 central time-slices for
both  $\vec{p}=(0,0,0)$ and $(0,0,1)$ correlators,  as  illustrated in
Figs. 5 and 6.  The  behavior of individual  $B_\CO$ terms is  very
similar to the quenched numbers shown in Tables 2a--3b.

Before  presenting  the final results, we   first discuss a  technical
drawback of this calculation.  On these lattices only Wuppertal source
correlators are available, so with our  method some contamination from
higher  momentum kaon states  is present  in  the  data.  For momentum
transfer $\vec{p}=(0,0,0)$, the  largest contamination  comes from the
propagation of   $\vec{p}=(0,0,1)$ kaons   across   the lattice.  (The
contamination in the $\vec{p}=(0,0,0) \to (0,0,1)$ data comes from the
presence of $\vec{p}= (0,0,1) \to  (0,0,2)$ terms.)  This contribution
is suppressed  by two factors: the exponential  suppression due to the
extra energy  of the $\vec{p}=(0,0,1)$ state,  and  the square  of the
ratio  of amplitudes  for creating  a $\vec{p}=(0,0,1)$  kaon versus a
$\vec{p}=(0,0,0)$  kaon   by  the  Wuppertal  source.   These  factors
increase as  the kaon mass decreases.   They are  similar  on our four
sets of lattices.    At $\beta=5.6$ and $\kappa=0.157$, our  estimates
are $\approx 10$  and $(1.5)^2$.  There is  also an enhancement factor
because the  matrix element   between higher  momentum  kaon states is
larger.  We estimate this factor using VSA to be $(E(p=1) / M_K)^2
\approx (1.5)^2$.  These three factors combine  to increase the result
for \bk\ by roughly $10\%$.  We note that in the case  of the quenched
lattices,  having $40$ time-slices reduced this   contamination to the
level of a few percent.   This is evident  on  comparing the $S-S$ and
the $S-W$ or the $W-W$ results in Tables 2a-3b.

We again calculate $B_K^L$ for three values of the effective coupling:
$0.0$, $g^2$ and $1.75 g^2$.  The lattice scale on individual lattices
(without extrapolation to the chiral  limit)  is not well defined, and
we  simply set $\mu a  = 1$.  The results  are  shown  in Table 6.  As
explained  above, the results for  \bk\ are  likely  to be $\sim 10\%$
larger  due to  contamination from  higher  momentum kaon states.   In
addition, \bk\ has to be extrapolated to the physical kaon mass.
%% Also the higher order terms in Eq. \bkdiff\ may be large!!!
Thus,  the only conclusion  we  can  draw is  that   the  quenched and
dynamical results are  in qualitative agreement  for quarks masses  in
the range $m_s < m_q < 3 m_s$.

\newsec{$B$-parameter for the Left-Right electromagnetic penguins}

There are two additional 4-fermion operators that we analyze using the
data in Tables  2a-3b.  These  are the $\Delta  I =  3/2$  part of the
left-right  electromagnetic   penguin operators $\CO_7$  and  $\CO_8$.
They alone contribute to the imaginary  part of the  $I = 2$ amplitude
and therefore   give the    dominant electromagnetic   contribution to
$\epsilon'/\epsilon$.   A  knowledge    of  their  $B$-parameters   is
phenomenologically important as discussed in Ref.~\ref
\lusignoli{
  M. Lusignoli, L. Maiani, G. Martinelli and L. Reina,
  \NPB{369} (1992) 139.
}.
Taking just the $\Delta I = 3/2 $ part of the operators simplifies the
numerical calculation as the ``eye" contractions cancel in  the flavor
SU(2) limit.

In principle  one would like to  calculate the matrix elements  of the
penguin operators
\eqn\Oseven{
  \CO_7 =     (\bar s_a \gamma_\mu L d_a) \  \big[
                 (\bar u_b \gamma_\mu R u_b) - {1 \over 2}
                 (\bar d_b \gamma_\mu R d_b) - {1 \over 2}
                 (\bar s_b \gamma_\mu R s_b) \big],
}
\eqn\Oeight{
  \CO_8 =     (\bar s_a \gamma_\mu L d_b) \  \big[
                 (\bar u_b \gamma_\mu R u_a) - {1 \over 2}
                 (\bar d_b \gamma_\mu R d_a) - {1 \over 2}
                 (\bar s_b \gamma_\mu R s_a) \big],
}
between a $K^+$ and a $\pi^+$.
Instead, we calculate the $\Delta I= 3/2$ part given by the operators
\eqn\LRseven{
  \CO_7^{3/2}
  = (\bar s_a \gamma_\mu L d_a) \  \big[
    (\bar u_b \gamma_\mu R u_b) -
    (\bar d_b \gamma_\mu R d_b) \  \big] +
   (\bar s_a \gamma_\mu L u_a) (\bar u_b \gamma_\mu R d_b),
}
\eqn\LReight{
  \CO_8^{3/2}
  = (\bar s_a \gamma_\mu L d_b) \  \big[
    (\bar u_b \gamma_\mu R u_a) -
    (\bar d_b \gamma_\mu R d_a) \  \big] +
    (\bar s_a \gamma_\mu L u_b) (\bar u_b \gamma_\mu R d_a).
}
Note  that  the overall normalization is unimportant  as it cancels in
the  $B$-parameters.   The 1-loop perturbatively corrected versions of
these operators have been calculated  in Refs.  \pertmarti\ \pertucla,
and  are linear  combinations   of  the operators  labeled $\CO_1$ and
$\CO_2$ therein.  The   matrix elements of  these corrected  operators
between a $K^+$ and a $\pi^+$ are, in the flavor SU(2) limit,
\eqn\LRsevenME{\eqalign{
 \CM_7(\vec p)
  = &{ \bigg(1 + {g^2 \over {16\pi^2}} Z_1(r,a\mu) \bigg) }
       \bigg[ 2\CP^1  -  2\CS^1 + \CV^2  -  \CA^2 \bigg] \cr
       + &  {{ g^2 \over {16\pi^2}}\,{{(Z_2-Z_1)} \over 3}}
                 \bigg[ 2\CP^2 - 2 \CS^2 + \CV^1 -  \CA^1  \bigg] \cr
       + &  {{g^2 \over {16\pi^2}}\,{{r^2 Z^*(r) } \over 12}}
                 \bigg[ 4\CP^1 - 12 \CP^2 + 16\CS^1 - 16 \CS^2 \cr
         & \qquad\qquad\qquad
       - 7 \CV^1 - 11 \CV^2 - 9 \CA^1 - 5\CA^2 - 6 \CT^1
       + 2 \CT^2  \bigg]. \cr
}}
and
\eqn\LRsevenME{\eqalign{
 \CM_8(\vec p)
  = &{ \bigg(1 + {g^2 \over {16\pi^2}} Z_2(r,a\mu) \bigg) }
        \bigg[ 2\CP^2  - 2\CS^2 + \CV^1  -  \CA^1 \bigg] \cr
        + &  {{g^2 \over {16\pi^2}}\,{{r^2 Z^*(r) } \over 12} }
                   \bigg[ 18 \CP^1 - 38 \CP^2 + 14\CS^1 - 26\CS^2 \cr
          & \qquad\qquad\qquad
            -5\CV^1 - \CV^2 + \CA^1 - 3 \CA^2 - 16 \CT^1  \bigg]. \cr
}}
where, if necessary,  we have  made  a  spin Fierz  transformation  to
recast  all the  terms as  two-spinor   loops.  The  corresponding VSA
contractions are
\eqn\VSAsevenME{
 \CM^{VSA}_7(\vec p)
  = \bigg[ {2 \over 3} Z_P^2\,\CP^2  -  Z_A^2\,\CA^2 \bigg] ,
}
and
\eqn\VSAsevenME{
 \CM^{VSA}_8(\vec p)
  = \bigg[ 2 Z_P^2\,\CP^2  -  {1 \over 3} Z_A^2\,\CA^2 \bigg] .
}
The $B$-parameters are the ratios of, for example, the  matrix element
of  $O_7$ to its VSA.   We evaluate these  in  the  SU(3) limit,  \ie\
degenerate   $u$, $d$   and $s$  quarks.   The   1-loop values of  the
renormalization constants  for  these  LR operators are  also given in
Table 1.   Note that in the $DRED$  scheme the operator  $\CO_8^{3/2}$
does not  mix  with the  scheme dependent   operator  $\bar  {\CO}$ of
Ref.~\pertucla.  It  is for this reason  that we choose  this  scheme,
although, our analysis shows that the results   are only weakly scheme
dependent.

In the chiral limit  these matrix  elements are expected  to behave as
$c+dm_\pi^2+\ldots$.  There are $O(a)$ corrections in the coefficients
$c$ and  $d$  due to the lattice discretization.   At present the only
way to reduce these is to use an improved action and/or work at weaker
coupling.  In this study   we  do not   have  any control   over these
corrections and we simply give the lattice results for the Wilson action.

The quality of the signal is shown in Figs. 7 and 8.   In the analysis
of these  LR operators we  find that using  the full covariance matrix
produced estimates that are about $1 \sigma$ lower than the fit values
shown unless we  significantly decrease the range  of  the  fit.   The
error estimates with and without  using the full  covariance matrix in
the fits are  essentially the same.  This indicates  that the data  at
different time  slices is highly  correlated and larger  statistics is
needed to  reliably include the correlations.  We   choose  to use the
full range  of the  plateau  in the  fit  and quote  results  obtained
without including the correlations.

As in the case of the  LL operator, in order to  quote a value for the
$B$-parameters we  have to specify the value  of $\gsqeff$ and $\mu a$
used  in the  perturbatively improved operators.  In  Table 7 we quote
results for a  number of  choices in order to  give an estimate of the
sensitivity of the results to variation in these parameters.  The data
show that this could  be a $ 10\%$ effect, so it is important to make
a good choice of $\gsqeff$.

The data also show a small increase in the $B$-parameters as the quark
mass is decreased.  Linearly  extrapolating the $\gsqeff = 1.75$, $\mu
a = \pi$ results to the physical kaon mass, our best estimates are
\eqn\finalnumbers{\eqalign{
  B_7^{3/2} &=  0.89(4),  \cr
  B_8^{3/2} &=  0.93(5).  \cr
}}
These values are slightly smaller than the numbers used by Lusignoli
\etal\  \lusignoli\ in their analysis of $\epsilon'  / \epsilon$; they
used  $B^{3/2}_7   = B^{3/2}_8 = 1.0 \pm   0.1$.  To   make a complete
determination of $\epsilon'/\epsilon$ we  need to calculate many other
matrix elements, for example of  the strong penguin operators  $\CO_5$
and  $\CO_6$, for which the  lattice technology  is  still unreliable.
For  this reason   we do  not   consider it  opportune to repeat   the
phenomenological analysis.

We can  make a direct  comparison of lattice results for $\CO_7^{3/2}$
with those obtained  by  Bernard  and Soni  on the subset  of lattices
described in the analysis   of $B_K$.   Their result,  obtained  using
$\gsqeff=1.338$ and  $\mu a = 1.7  \pi$ in  the $DR({\overline {EZ}})$
scheme, is 0.965(41) at $\kappa=0.155$, to be  compared with 0.971(50)
obtained by us.  This comparison suggests that for the matrix elements
of  LR operators,  our method of  sandwiching   the  operator  between
smeared  sources is no  better than using   propagators from  a single
source point.  On the other hand the fact that smeared sources yield a
plateau over a large range of time-slices gives reassurance that one
potential source of systematic error is under control.

\newsec{Conclusions}

We show that the   calculation of the kaon $B$-parameters  with Wilson
fermions  is  significantly  improved  by the   use of non-local quark
sources.   By   using a    combination of Wuppertal   and  wall source
correlators, we demonstrate  that the on-shell  matrix elements can be
calculated at   non-zero  momenta.

By combining results at $\vec p=(0,0,0)$ and $(0,0,1)$, we carry out a
non-perturbative   subtraction  of    the  lattice    artifacts in the
calculation of  \bk.  Even though  we  cannot  take  into account  the
artifact $\gamma$,  our  results are  in  good  agreement   with those
obtained  with staggered fermions.   On the  basis of this exploratory
study we feel confident that the momentum subtraction procedure indeed
works.  To  make further improvements  and reduce the $O(a)$ artifacts
one needs to repeat the calculations with an improved lattice action and
on a larger physical lattice with smaller $\vec p_{min}$.

We find a clean plateau in the  data for the  $B$-parameters of the LR
electromagnetic penguin operators.  The results   show that VSA  works
much    better  for these   operators.   All  the   B-parameters  vary
significantly with the choice of $\gsqeff$   used in  the perturbative
renormalization coefficients.   Our  final  estimates are given  using
the  value advocated by Lepage  and Mackenzie in Ref.~\effgsq,
$i.e.$ $\gsqeff = 1.75$.

The method of using the combination of Wuppertal and  wall correlators
can   be  extended to study other  3-point   correlation functions, in
particular structure  functions  and form   factors.  This work  is in
progress.

\bigskip
\centerline{\bf Acknowledgments}

We thank C. Bernard  and G. Martinelli for  providing unpublished data
and for discussions.   The  $16^3\times40$ lattices were  generated at
NERSC at Livermore using a  DOE allocation.   The calculation of quark
propagators   and  the  analysis  has  been   done at  the  Pittsburgh
Supercomputing Center, San  Diego Supercomputer Center,  NERSC and Los
Alamos  National Laboratory.  We  are very  grateful to Jeff  Mandula,
Norm  Morse, Ralph Roskies, Charlie Slocomb  and  Andy White for their
support of this project.  This  research was supported  in part by the
National Science Foundation under Grant No.  PHY89-04035.  RG, GWK, AP
and  SRS thank Institute for Theoretical  Physics,  Santa Barbara  for
hospitality during  part of this  work.   AP  also  thanks  Los Alamos
National Laboratory  for hospitality during  the course of this  work.
GWK is supported in  part by  DOE Outstanding Junior  Investigator and
NSF Presidential Young  Investigator  programs.   SRS is supported  in
part  by DOE  contract  DE-AC05-84ER40150  and  by  an Alfred P. Sloan
Fellowship.

\vfill\eject

% Figures %%%%%%%%%%%%%%%%%%%%%%%%%%%%%%%%%%%%%%%%%%%%%%%%%%%%%%%%%%%%

\input epsf
\def\inputfigure#1#2{
  \pageinsert\parindent=0pt %%%\baselineskip=1.5\normalbaselineskip
  $$\epsfysize=\hsbody\epsfbox{#1}$$
  \par\bigskip\bigskip\bigskip #2 \vfill\endinsert}

\inputfigure{ 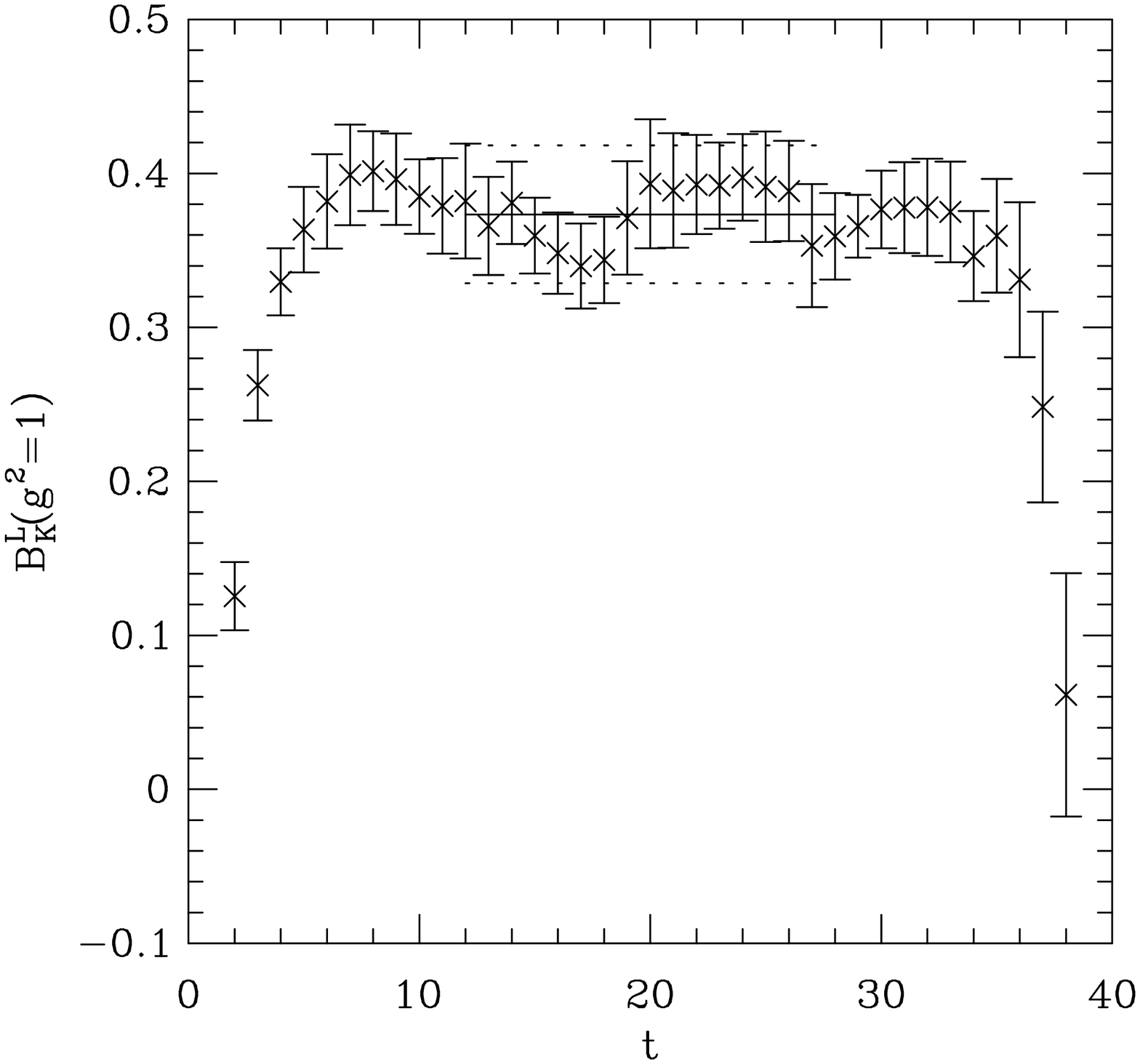 }{
  {\bf Fig.   1a.} The ratio  of correlators for  the  lattice parameter
  $B_K^L (g^2=1,\  \mu a =  1.0)$  at  $\kappa=0.154$  and  for momentum
  transfer $\vec p   =(0,0,0)$.   The data  are  obtained using operator
  $\gamma_5$ as the kaon source.}

\inputfigure{ 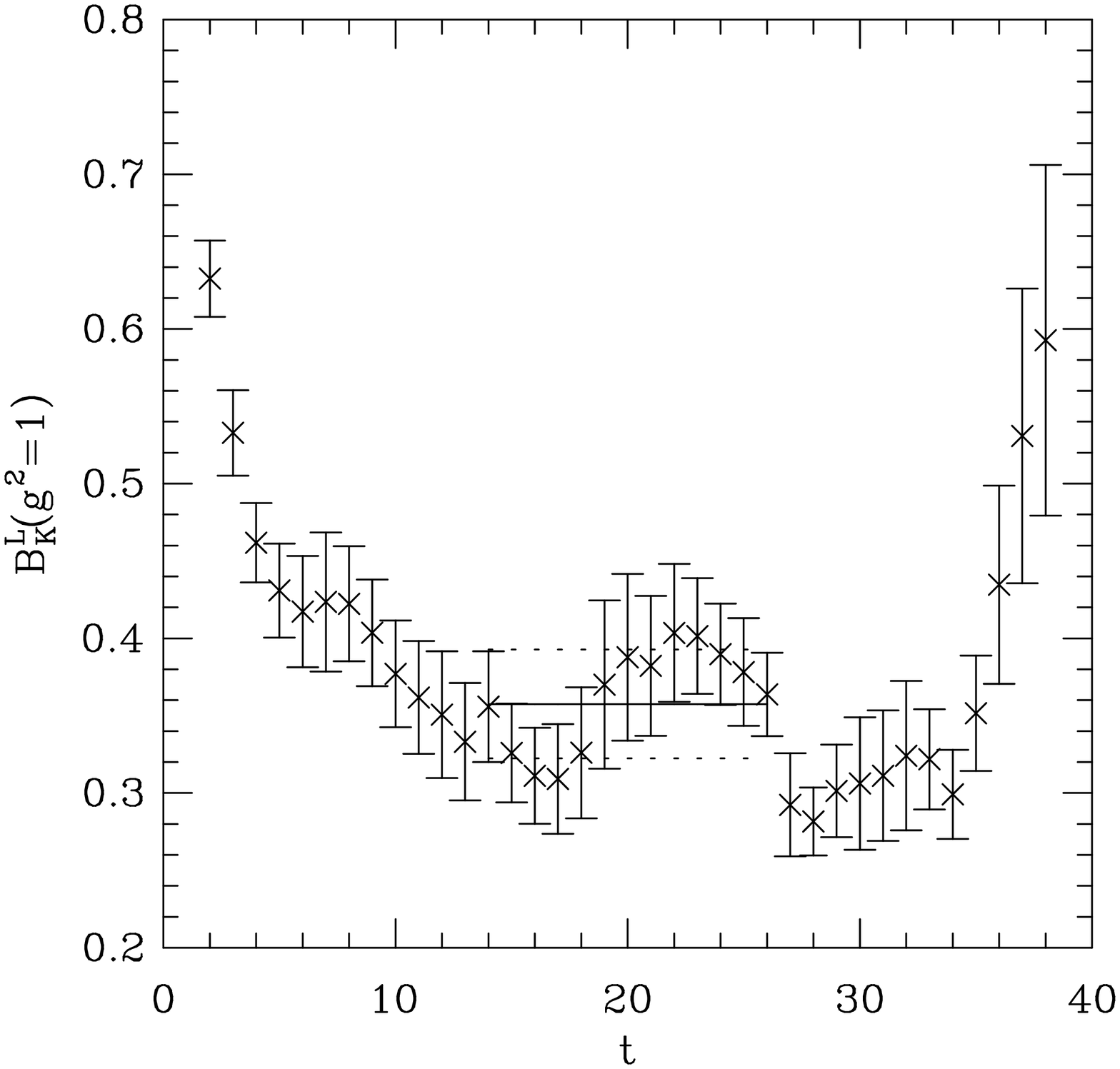 }{
  {\bf Fig. 1b.} The same as in Fig. 1a,  but using operator  $A_4$ as
  the kaon source.}

\inputfigure{ 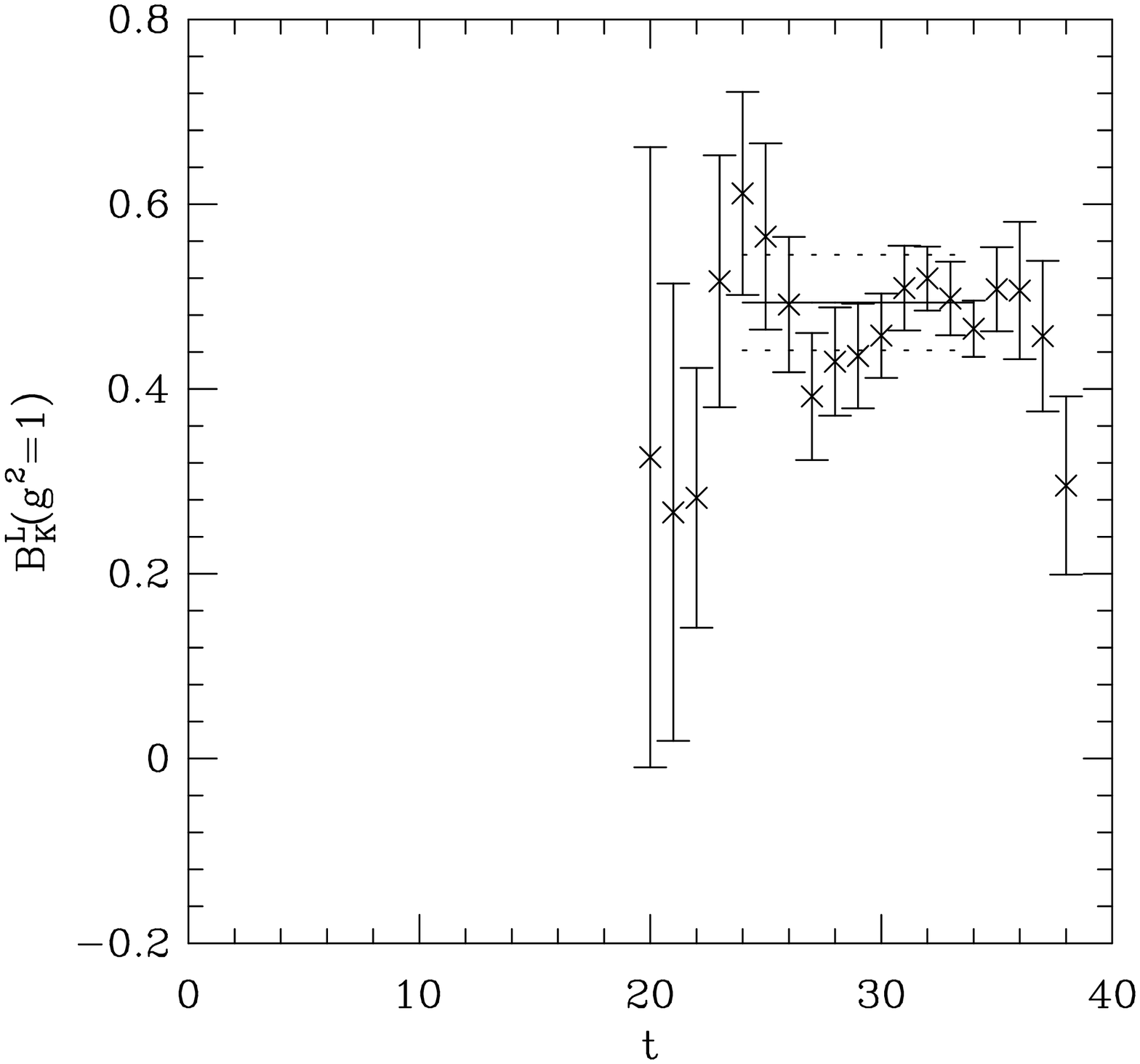 }{
  {\bf  Fig. 2.} The same  as in Fig. 1a, but   for $\kappa=0.154$ and
  $\vec p =(0,0,1)$.}

\inputfigure{ 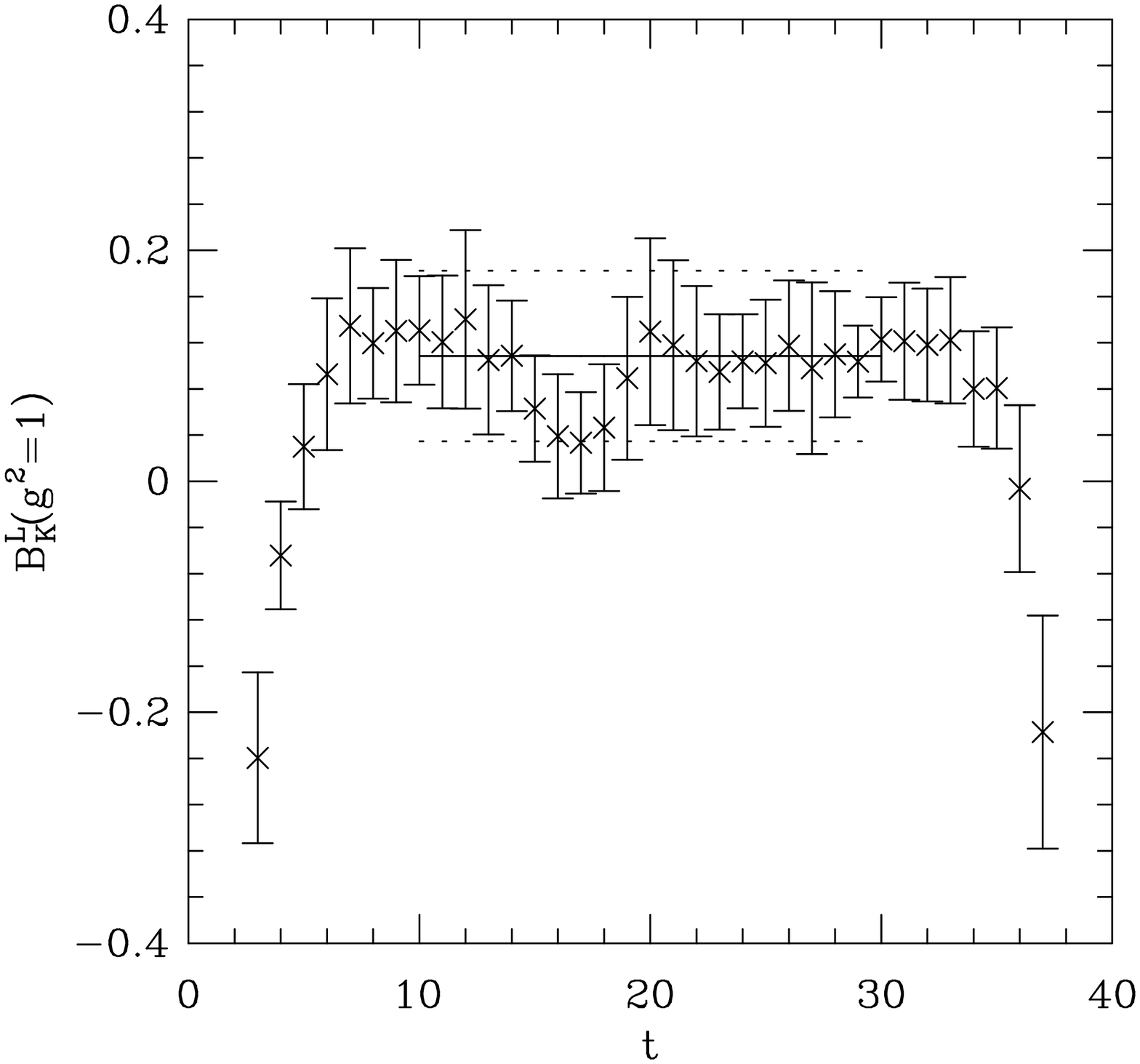 }{
  {\bf Fig.  3.} The same as  in Fig. 1a,  but  for $\kappa=0.155$ and
  $\vec p =(0,0,0)$.}

\inputfigure{ 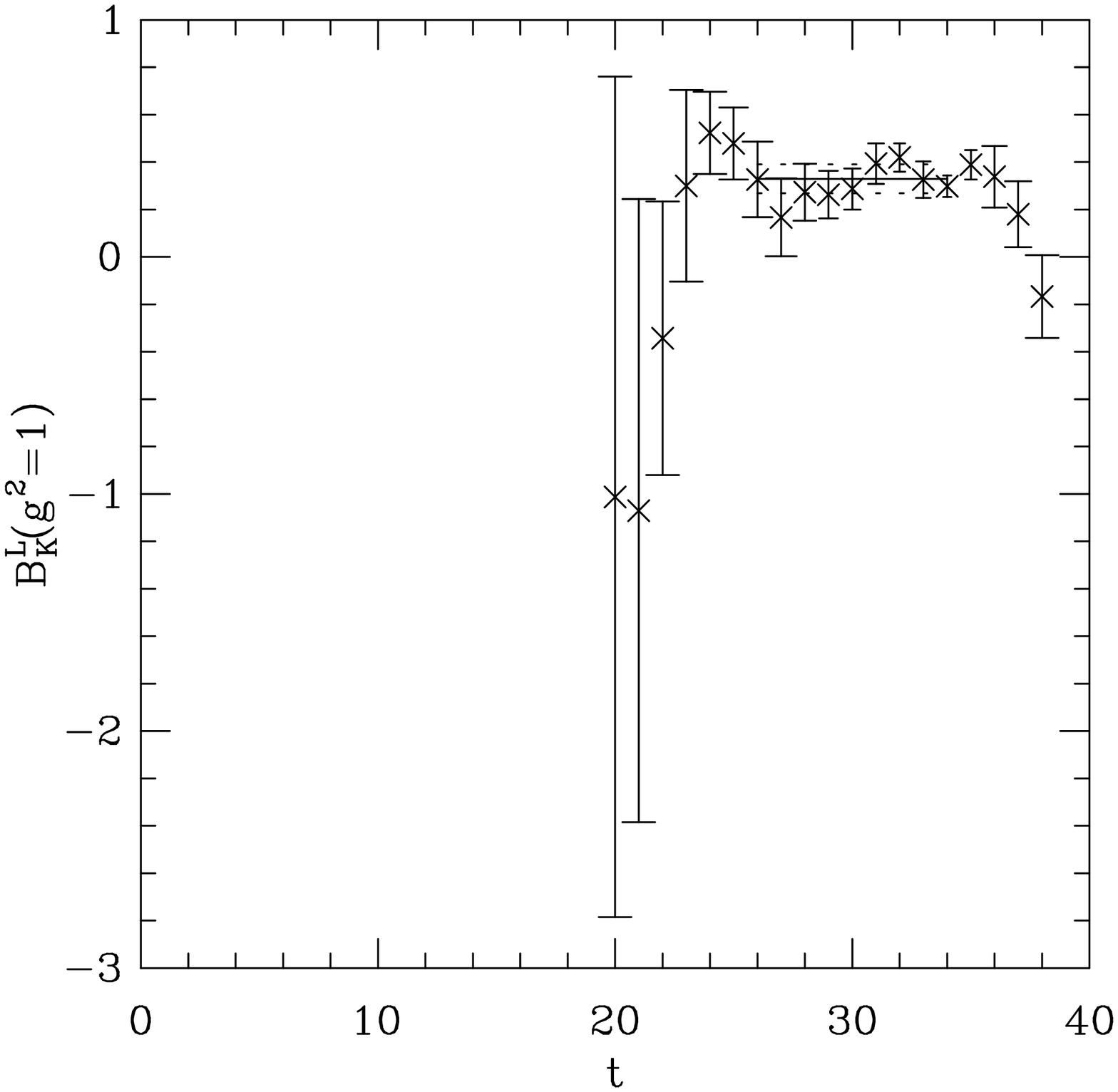 }{
  {\bf  Fig. 4.} The same  as in Fig.  1a,  but for $\kappa=0.155$ and
  $\vec p =(0,0,1)$.}

\inputfigure{ 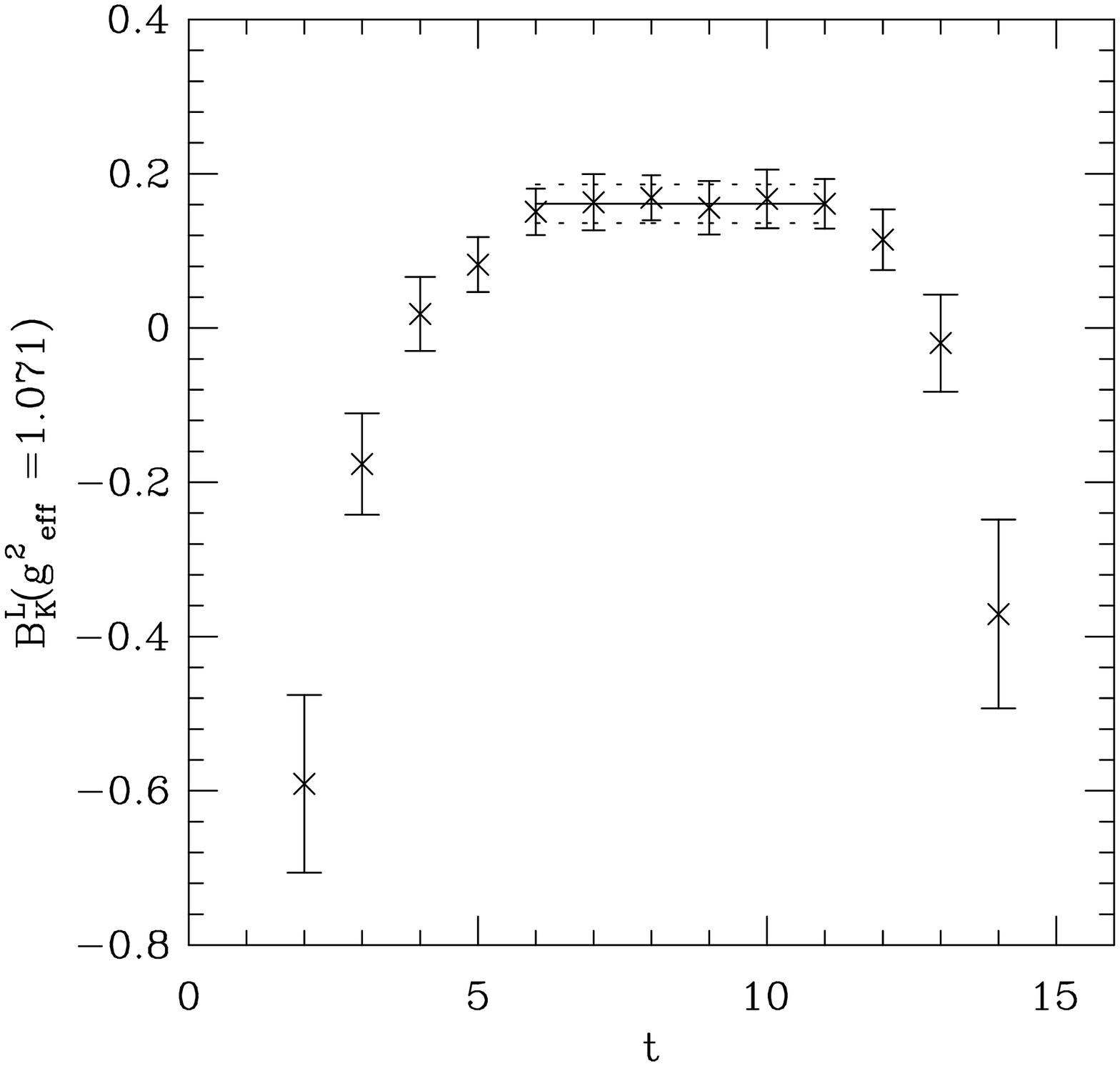 }{
  {\bf Fig. 5.} The ratio  of correlators for   the  lattice parameter
  $B_K^L(\gsqeff=1.071,\ \mu a = 1)$ measured  on $16^4$ lattices generated
  with two degenerate flavors of dynamical Wilson fermions. The lattice
  parameters   are $\beta = 5.6$, $\kappa=0.157$ and momentum transfer
  $\vec p =(0,0,0)$, and we use $\bar s\gamma_5 d$ as the kaon source.}

\inputfigure{ 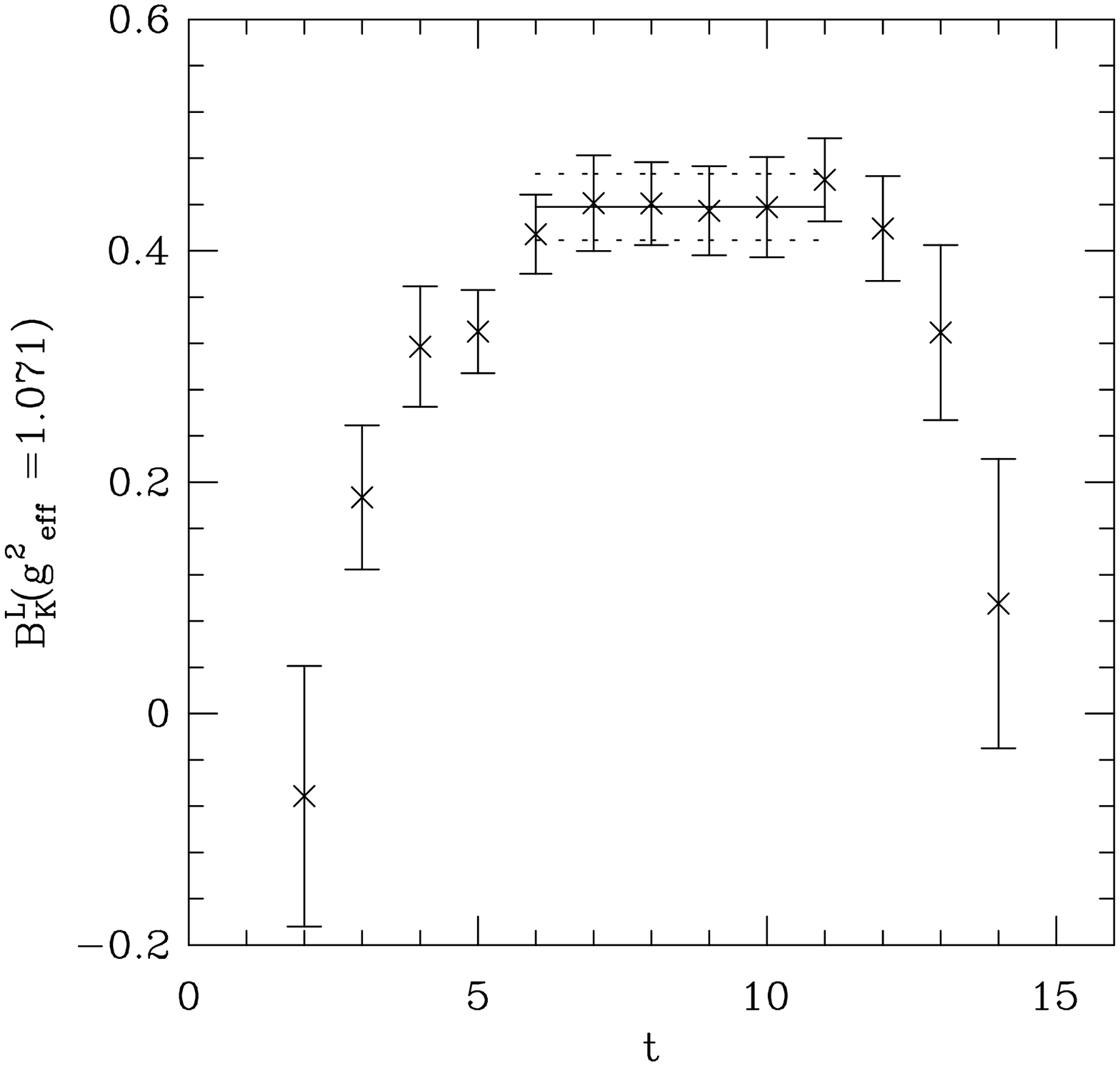 }{
  {\bf Fig. 6} The same as in Fig. 5, but for $\vec p =(0,0,1)$.}

\inputfigure{ 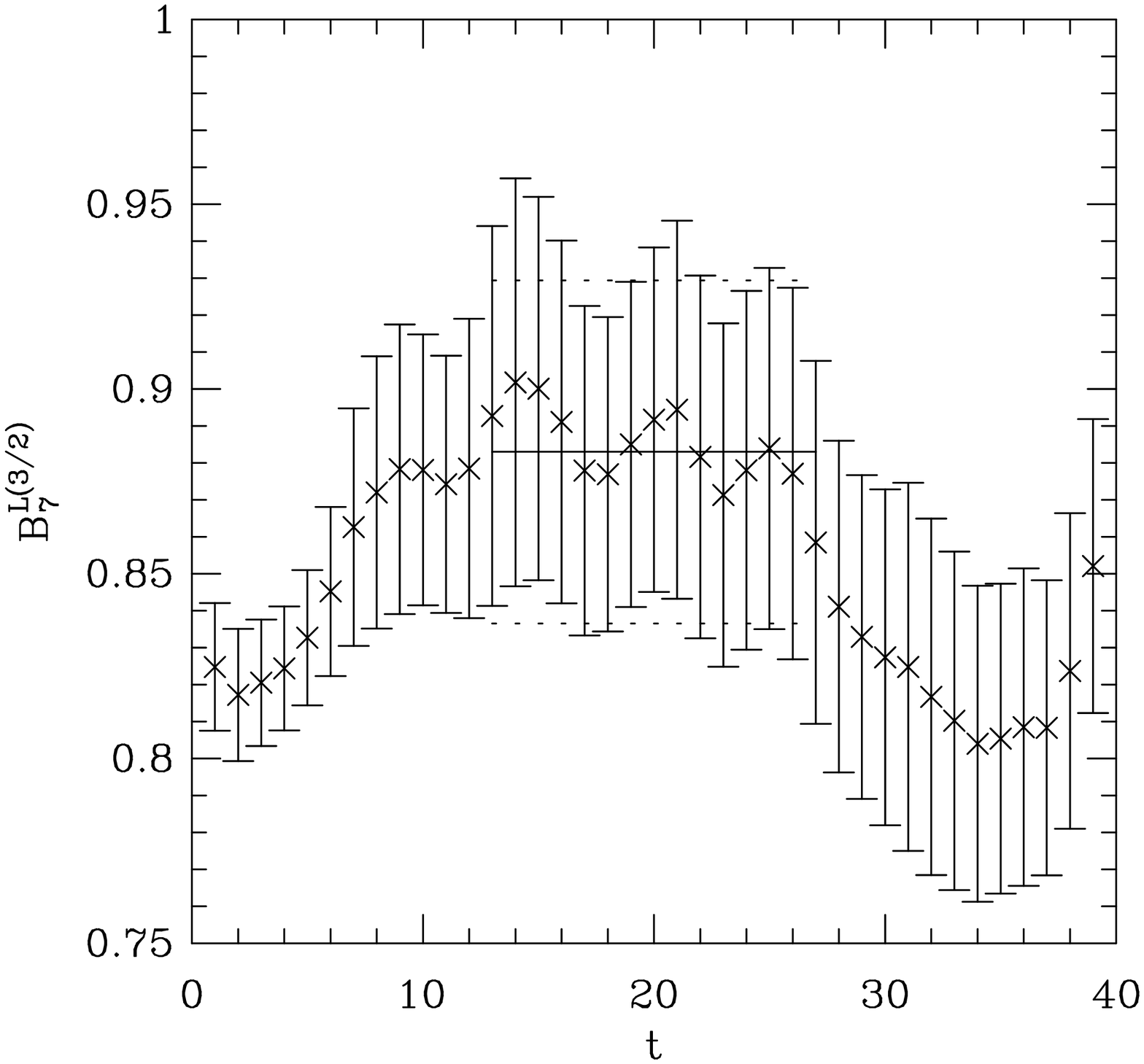 }{
  {\bf Fig. 7} $B$-parameter for the LR electromagnetic penguin
  operator $\CO_7^{3/2}$.  The data are obtained using $\gsqeff = 1.75$
  and $\mu a = \pi$. }

\inputfigure{ 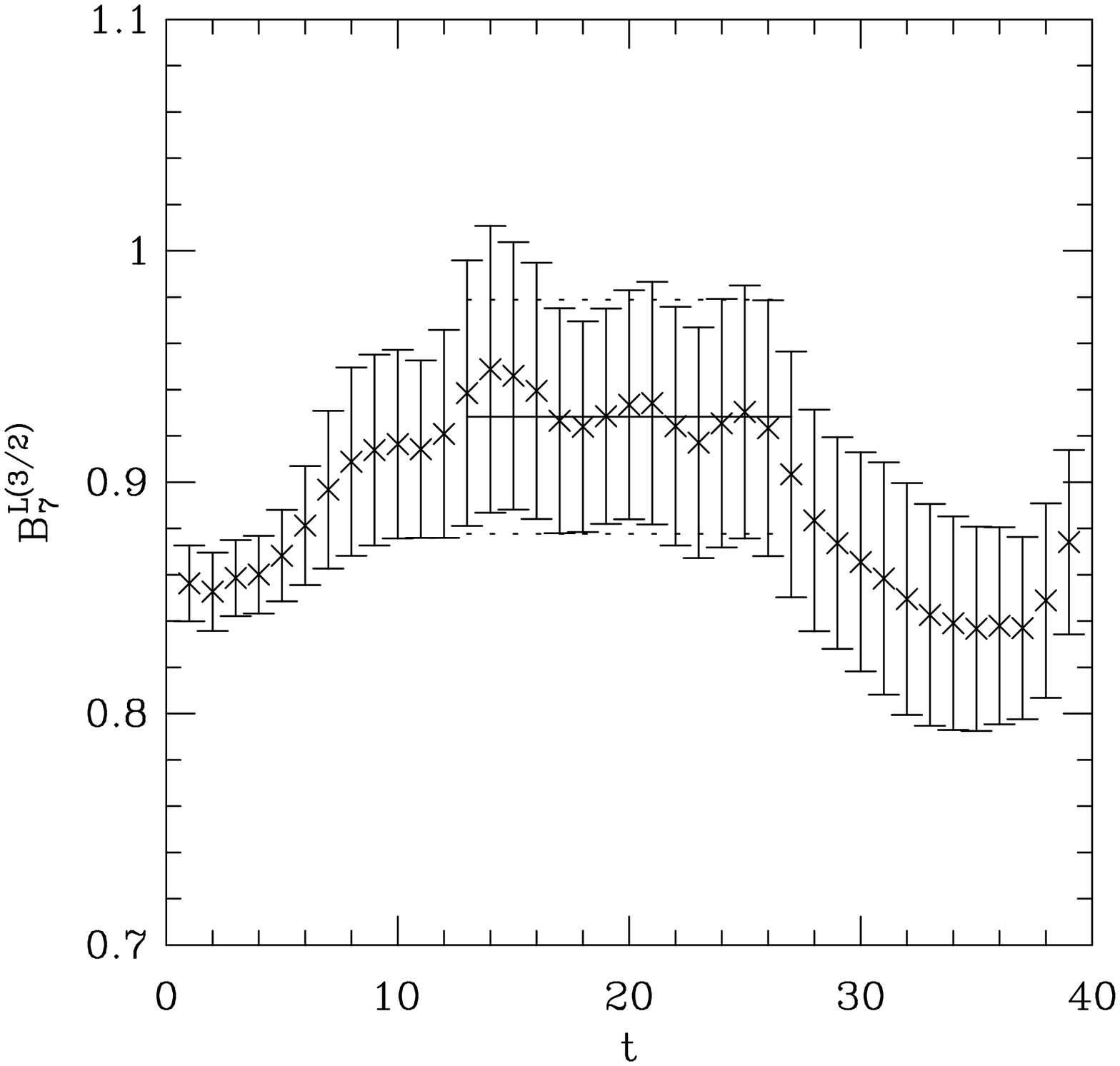 }{
  {\bf Fig. 8} $B$-parameter for the LR electromagnetic penguin
  operator $\CO_8^{3/2}$.  The data are obtained using $\gsqeff = 1.75$
  and $\mu a = \pi$. }

\bigskip
\vfill\eject

% Tables %%%%%%%%%%%%%%%%%%%%%%%%%%%%%%%%%%%%%%%%%%%%%%%%%%%%%%%%%%%%%

  \pageinsert\parindent=0pt %%%\baselineskip=1.5\normalbaselineskip
  %% t1_zfac.tex
$$
\vbox{\hbox{\indent\vbox{\tabskip=0pt\offinterlineskip
\def\tlr{\noalign{\hrule}}
\def\q{\quad}
\def\s{\phantom{-}}
\def\sq{\phantom{-}\q}
\def\MSbar{{NDR} }
\def\dredMS{DRED}
\def\dredEZ{DR({\overline {EZ}}) }

\def\mua{{\rm log}(\mu a) }
\def\lam{\lambda}

\halign {\strut#& \vrule#\tabskip=3pt&
  \hfil$#$\hfil&\vrule\vrule#&
  \hfil$#$\hfil&\vrule\vrule#&
  \hfil$#$\hfil&\vrule#&
  \hfil$#$\hfil&\vrule#\tabskip=0pt\cr\tlr
\omit&height2pt&\omit&&\omit&&\omit&&\omit&\cr
&&     && \dredEZ                       &&  \dredMS       &&
 \MSbar          &\cr
\omit&height2pt&\omit&&\omit&&\omit&&\omit&\cr\tlr
\omit&height2pt&\omit&&\omit&&\omit&&\omit&\cr\tlr
\omit&height2pt&\omit&&\omit&&\omit&&\omit&\cr
&& Z_A && 1 - 15.796 C_F \lam           &&\s 0.5C_F \lam  &&
\sq 0            &\cr
\omit&height2pt&\omit&&\omit&&\omit&&\omit&\cr\tlr
\omit&height2pt&\omit&&\omit&&\omit&&\omit&\cr
&& Z_P && 1 + C_F \lam (6\mua - 21.596) &&\s 0            &&
\sq -C_F \lam    &\cr
\omit&height2pt&\omit&&\omit&&\omit&&\omit&\cr\tlr
\omit&height2pt&\omit&&\omit&&\omit&&\omit&\cr
&& Z_+ && -50.174 - 4\mua               &&\s {7 \over 3}  &&
\sq {14 \over 3} &\cr
\omit&height2pt&\omit&&\omit&&\omit&&\omit&\cr\tlr
\omit&height2pt&\omit&&\omit&&\omit&&\omit&\cr
&& Z_- && -45.308 + 8\mua               &&\s -{2 \over 3} &&
\sq -{4 \over 3} &\cr
\omit&height2pt&\omit&&\omit&&\omit&&\omit&\cr\tlr
\omit&height2pt&\omit&&\omit&&\omit&&\omit&\cr
&& Z_1 && -49.364 - 2 \mua              &&\s {11 \over 6} &&
\sq {11 \over 3} &\cr
\omit&height2pt&\omit&&\omit&&\omit&&\omit&\cr\tlr
\omit&height2pt&\omit&&\omit&&\omit&&\omit&\cr
&& Z_2 && -42.064 + 16 \mua             &&\s -{8 \over 3} &&
\sq -{16 \over 3}&\cr
\omit&height2pt&\omit&&\omit&&\omit&&\omit&\cr\tlr
\omit&height2pt&\omit&&\omit&&\omit&&\omit&\cr
&& Z^* && 9.6431                        &&\s  0           &&
\sq 0            &\cr
\omit&height2pt&\omit&&\omit&&\omit&&\omit&\cr\tlr
}}}}$$

%%
%%  ADD THE 3 AND 4 COLUMNS TO 2 TO GET NUMBERS IN THOSE SCHEMES!!!!
%%

%%%%%%%%%%%%%%%%%%%%%%%%%%%%%%%%%%%%%%%%%%%%%%%%%%%%%%%%%%%%%%%%%%%%%%%%%
%% NOTES
%%%%%%%%%%%%%%%%%%%%%%%%%%%%%%%%%%%%%%%%%%%%%%%%%%%%%%%%%%%%%%%%%%%%%%%%%
%% BSD (\delta=1/4)                     -------->   Guido
%% 16K_2                                    =        B-2
%% \Delta_{\Sigma_1} - 4\delta              =        \Delta_{\Sigma_1}
%%
%% compare BSD Eq. 4.28 with Guido Eq. 12
%%%%%%%%%%%%%%%%%%%%%%%%%%%%%%%%%%%%%%%%%%%%%%%%%%%%%%%%%%%%%%%%%%%%%%%%%
%%
%% Guido                                --------->   MS \bar
%%  \Delta_{S,P}       +2               --------->    \Delta_{S,P}
%%  \Delta_{V,A}       +0.5             --------->    \Delta_{V,A}
%%  \Delta_{\Sigma_1}  -1               --------->    \Delta_{\Sigma_1}

%% The first two lead to the change B ---> B+2
%% (to go from guido's formulae to MS-bar as per Guido's Eq. 5).
%%%%%%%%%%%%%%%%%%%%%%%%%%%%%%%%%%%%%%%%%%%%%%%%%%%%%%%%%%%%%%%%%%%%%%%%%
%% Thus simple substitution rule is take Guido's Eq. 12 and where ever
%% there is B and \Delta_{\Sigma_1}  add to them the following:
%%
%%  BSD                           Guido                  MS \bar
%%  -2                              B                     + 2
%%  - 4\delta                     \Delta_{\Sigma_1}       + 1
%%
%%  where transformation to BSD only apply to 4-fermi operators!
%%  between BSD and Guidoonly apply to 4-fermi operators!

  \par 
  {\bf Table  1.}  Summary of 1-loop perturbative results for the
  various $Z$ factors needed in our calculations in three different
  continuum regularization schemes.  The two constants are $C_F = 4/3$
  and $\lambda = g^2 /(16 \pi ^2)$.  The results in $DRED$ and $NDR$
  schemes are the sum of those in $DR(\overline{EZ})$ and the entries
  in their respective columns.    These expressions are extracted from
  Refs.~\pertmarti\ \pertucla\ and \ushmwilson.  The numerical results
  are taken from Ref.~\pertucla.  In the text all results are given in
  the $DRED$ scheme used in Ref.~\pertmarti, except when we compare raw
  lattice numbers against those in Ref.~\bsprivate. \vfill\endinsert

  \pageinsert\parindent=0pt %%%\baselineskip=1.5\normalbaselineskip
  %% t2a_54_all.tex
$$
\ninepoint
\vbox{\hbox{\indent\vbox{
\tabskip=0pt\offinterlineskip
\def\tlr{\noalign{\hrule}}
\def\CO{{\cal O}}
\halign {\strut#& \vrule#\tabskip=2pt&
  \hfil$#$\hfil&\vrule\vrule#&
  \hfil$#$\hfil&\vrule#&
  \hfil$#$\hfil&\vrule#&
  \hfil$#$\hfil&\vrule#&
  \hfil$#$\hfil&\vrule\vrule#&
  \hfil$#$\hfil&\vrule#&
  \hfil$#$\hfil&\vrule#&
  \hfil$#$\hfil&\vrule#&
  \hfil$#$\hfil&\vrule#\tabskip=0pt\cr\tlr
\omit&height2pt&\omit&&\omit&&\omit&&\omit&&
\omit&&\omit&&\omit&&\omit&&\omit&\cr
&&
&&PS[0]SP
&&PW[0]WP
&&PS[0]WP
&&PS[1]WP
&&AS[0]SA
&&AW[0]WA
&&AS[0]WA
&&AS[1]WA
& \cr\tlr
\omit&height2pt&\omit&&\omit&&\omit&&\omit&&
\omit&&\omit&&\omit&&\omit&&\omit&\cr\tlr
&&
&&  1.6
&&  1.1
&&  0.3
&&  0.8
&&  1.4
&&  1.2
&&  0.6
&&  0.5
  &\cr
&&
&& 14-26
&& 14-26
&& 14-26
&& 23-31
&& 12-28
&& 14-26
&& 10-30
&& 22-31
  &\cr
&& {\cal P}^1
&&   -4.30( 21)
&&   -4.03( 17)
&&   -3.99( 19)
&&   -2.58( 28)
&&   -3.86( 35)
&&   -3.86( 21)
&&   -3.70( 20)
&&   -2.45( 27)
  &\cr\tlr
&&
&&  2.3
&&  4.7
&&  2.0
&&  2.0
&&  7.7
&&  4.0
&&  2.8
&&  1.8
  &\cr
&&
&& 10-30
&& 12-28
&& 12-28
&& 24-33
&& 10-30
&& 11-29
&& 10-30
&& 22-31
  &\cr
&& {\cal S}^1
&&   -0.45(  4)
&&   -0.42(  6)
&&   -0.42(  3)
&&   -0.26(  5)
&&   -0.38(  8)
&&   -0.40(  6)
&&   -0.38(  4)
&&   -0.26(  5)
  &\cr\tlr
&&
&&  2.2
&&  6.0
&&  4.3
&&  1.2
&&  4.0
&&  6.1
&&  3.2
&&  1.8
  &\cr
&&
&& 16-28
&& 12-28
&& 13-27
&& 26-33
&& 10-30
&& 10-30
&& 10-30
&& 22-32
  &\cr
&& {\cal V}_s^1
&&    0.40( 18)
&&    0.35( 10)
&&    0.35(  8)
&&    0.10( 11)
&&    0.36(  9)
&&    0.39( 11)
&&    0.31(  6)
&&    0.10( 10)
  &\cr\tlr
&&
&&  4.3
&&  1.3
&&  1.1
&&  0.8
&&  2.3
&&  1.2
&&  2.7
&&  0.9
  &\cr
&&
&& 11-29
&& 12-28
&& 14-26
&& 26-33
&& 15-25
&& 14-26
&& 10-30
&& 22-31
  &\cr
&& {\cal V}_t^1
&&    0.24(  6)
&&    0.22(  2)
&&    0.21(  2)
&&    0.12(  2)
&&    0.22(  3)
&&    0.20(  2)
&&    0.22(  2)
&&    0.11(  3)
  &\cr\tlr
&&
&&  3.5
&&  2.0
&&  2.2
&&  0.2
&&  2.2
&&  2.7
&&  2.0
&&  0.4
  &\cr
&&
&& 12-28
&& 12-28
&& 11-29
&& 25-34
&& 10-30
&& 14-26
&& 10-30
&& 26-33
  &\cr
&& {\cal A}_s^1
&&   -0.60(  6)
&&   -0.56(  6)
&&   -0.56(  5)
&&   -0.33(  4)
&&   -0.54(  3)
&&   -0.50(  5)
&&   -0.52(  6)
&&   -0.28(  6)
  &\cr\tlr
&&
&&  3.8
&&  1.0
&&  4.2
&&  2.9
&&  0.8
&&  1.4
&&  0.6
&&  1.6
  &\cr
&&
&& 12-28
&& 14-26
&&  9-31
&& 27-33
&& 14-26
&& 14-26
&& 14-26
&& 27-35
  &\cr
&& {\cal A}_t^1
&&    0.11(  2)
&&    0.12(  2)
&&    0.10(  2)
&&    0.15(  3)
&&    0.10(  3)
&&    0.12(  2)
&&    0.09(  1)
&&    0.15(  4)
  &\cr\tlr
&&
&&  1.8
&&  1.5
&&  0.9
&&  1.9
&&  0.6
&&  2.8
&&  1.1
&&  0.7
  &\cr
&&
&& 10-30
&& 15-25
&& 10-30
&& 25-34
&& 16-24
&& 12-28
&& 10-30
&& 23-31
  &\cr
&& {\cal T}_s^1
&&    2.84( 21)
&&    2.63( 10)
&&    2.70( 10)
&&    1.59( 18)
&&    2.56( 34)
&&    2.59( 14)
&&    2.51( 16)
&&    1.54( 16)
  &\cr\tlr
&&
&&  3.3
&&  1.0
&&  0.7
&&  1.1
&&  1.5
&&  2.5
&&  0.6
&&  1.2
  &\cr
&&
&& 14-26
&& 15-25
&& 10-30
&& 25-32
&& 12-24
&& 14-26
&& 12-28
&& 24-31
  &\cr
&& {\cal T}_t^1
&&    2.93( 24)
&&    2.73( 10)
&&    2.79( 12)
&&    1.70( 21)
&&    2.67( 36)
&&    2.66( 17)
&&    2.58( 17)
&&    1.66( 20)
  &\cr\tlr
}}}}$$
  \par 
  {\bf Table   2a.} The one color  loop   contribution to  the lattice
  $B-$parameters for   individual operators  at  $\kappa=0.154$.  Each
  box shows the $\chi^2$  for a correlated fit, the  temporal range of
  the fit and the  fitted value.   The  space  and time components  of
  the operators have  been  shown separately.  The appropriate  ratios
  of  correlators  have  been calculated  using  two   different  kaon
  operators,  $\gamma_5$  and $A_4$,    and using  Wuppertal and  wall
  sources.  The notation, for  example, is: $PS[1]WP$  stands for  the
  four-fermion operator  with one unit  of lattice momentum sandwiched
  between Wuppertal and wall source kaons  each created  with operator
  $\gamma_5$.    All errors    are   calculated  using    the   single
  elimination jackknife method. \vfill\endinsert

  \pageinsert\parindent=0pt %%%\baselineskip=1.5\normalbaselineskip
  %% t2b_54_all.tex
$$
\ninepoint
\mkern -20mu
\vbox{\hbox{\indent\vbox{
\tabskip=0pt\offinterlineskip
\def\tlr{\noalign{\hrule}}
\def\CO{{\cal O}}
\halign {\strut#& \vrule#\tabskip=2pt&
  \hfil$#$\hfil&\vrule\vrule#&
  \hfil$#$\hfil&\vrule#&
  \hfil$#$\hfil&\vrule#&
  \hfil$#$\hfil&\vrule#&
  \hfil$#$\hfil&\vrule\vrule#&
  \hfil$#$\hfil&\vrule#&
  \hfil$#$\hfil&\vrule#&
  \hfil$#$\hfil&\vrule#&
  \hfil$#$\hfil&\vrule#\tabskip=0pt\cr\tlr
\omit&height2pt&\omit&&\omit&&\omit&&\omit&&
\omit&&\omit&&\omit&&\omit&&\omit&\cr
&&
&&PS[0]SP
&&PW[0]WP
&&PS[0]WP
&&PS[1]WP
&&AS[0]SA
&&AW[0]WA
&&AS[0]WA
&&AS[1]WA
& \cr\tlr
\omit&height2pt&\omit&&\omit&&\omit&&\omit&&
\omit&&\omit&&\omit&&\omit&&\omit&\cr\tlr
&&
&&  1.6
&&  0.9
&&  0.4
&&  0.8
&&  1.6
&&  1.1
&&  0.8
&&  0.6
  &\cr
&&
&& 12-28
&& 12-28
&& 10-30
&& 23-32
&& 12-28
&& 14-26
&&  9-31
&& 22-31
  &\cr
&& {\cal P}^2
&&  -12.09( 75)
&&  -11.60( 50)
&&  -11.26( 43)
&&   -7.15( 75)
&&  -10.99( 95)
&&  -11.05( 60)
&&  -10.77( 54)
&&   -6.81( 68)
  &\cr\tlr
&&
&&  2.9
&&  2.7
&&  2.8
&&  1.7
&&  2.7
&&  1.3
&&  2.0
&&  1.8
  &\cr
&&
&& 10-30
&& 14-26
&& 14-26
&& 25-33
&& 12-28
&& 14-26
&& 10-30
&& 22-31
  &\cr
&& {\cal S}^2
&&   -0.60( 10)
&&   -0.51(  9)
&&   -0.49(  6)
&&   -0.40(  8)
&&   -0.47( 14)
&&   -0.49(  6)
&&   -0.52(  7)
&&   -0.38(  7)
  &\cr\tlr
&&
&&  1.2
&&  1.3
&&  1.1
&&  0.6
&&  3.0
&&  2.5
&&  1.6
&&  1.6
  &\cr
&&
&& 15-25
&& 14-26
&& 12-28
&& 25-34
&& 10-30
&& 14-32
&& 10-32
&& 26-33
  &\cr
&& {\cal V}_s^2
&&    0.03(  3)
&&    0.04(  2)
&&    0.05(  2)
&&    0.01(  2)
&&    0.04(  3)
&&    0.04(  3)
&&    0.04(  2)
&&   -0.02(  3)
  &\cr\tlr
&&
&&  1.0
&&  2.1
&&  1.8
&&  1.3
&&  2.8
&&  2.1
&&  2.2
&&  1.5
  &\cr
&&
&& 14-26
&& 14-26
&& 10-30
&& 25-34
&& 12-28
&& 14-26
&& 10-30
&& 24-31
  &\cr
&& {\cal V}_t^2
&&    0.02(  0)
&&    0.02(  1)
&&    0.02(  0)
&&    0.01(  1)
&&    0.02(  0)
&&    0.02(  1)
&&    0.02(  0)
&&    0.02(  1)
  &\cr\tlr
&&
&&  5.1
&&  2.4
&&  1.9
&&  2.1
&&  2.4
&&  2.1
&&  1.9
&&  1.2
  &\cr
&&
&&  8-32
&& 11-26
&& 10-30
&& 25-34
&& 10-30
&& 14-26
&& 10-30
&& 25-33
  &\cr
&& {\cal A}_s^2
&&   -0.54(  5)
&&   -0.52(  4)
&&   -0.53(  4)
&&   -0.31(  3)
&&   -0.50(  7)
&&   -0.49(  4)
&&   -0.49(  5)
&&   -0.25(  4)
  &\cr\tlr
&&
&&  3.4
&&  4.2
&&  1.3
&&  0.8
&&  2.8
&&  3.3
&&  0.7
&&  0.8
  &\cr
&&
&& 13-24
&& 10-25
&& 15-25
&& 25-33
&& 12-24
&& 14-26
&& 16-24
&& 24-32
  &\cr
&& {\cal A}_t^2
&&    0.64(  4)
&&    0.62(  3)
&&    0.58(  2)
&&    0.64(  8)
&&    0.56(  5)
&&    0.58(  7)
&&    0.54(  3)
&&    0.60(  7)
  &\cr\tlr
&&
&&  5.3
&&  3.6
&&  3.2
&&  2.0
&&  3.6
&&  3.9
&&  3.0
&&  1.5
  &\cr
&&
&& 10-30
&& 12-28
&& 10-30
&& 28-34
&& 12-28
&& 14-26
&& 10-30
&& 23-35
  &\cr
&& {\cal T}_s^2
&&    0.05(  2)
&&    0.06(  2)
&&    0.05(  1)
&&    0.03(  1)
&&    0.05(  2)
&&    0.05(  1)
&&    0.04(  2)
&&    0.04(  1)
  &\cr\tlr
&&
&&  7.1
&&  1.5
&&  2.3
&&  1.0
&&  3.7
&&  2.1
&&  4.1
&&  1.1
  &\cr
&&
&& 14-26
&& 14-26
&& 10-30
&& 25-33
&&  8-31
&& 14-26
&&  6-34
&& 26-33
  &\cr
&& {\cal T}_t^2
&&    0.07(  2)
&&    0.05(  1)
&&    0.05(  1)
&&    0.03(  1)
&&    0.03(  5)
&&    0.04(  3)
&&    0.06(  2)
&&    0.05(  1)
  &\cr\tlr
}}}}$$
 \par 
  {\bf Table 2b.} The  same as Table  2a, but for the  two color
  loop contribution. \vfill\endinsert

  \pageinsert\parindent=0pt %%%\baselineskip=1.5\normalbaselineskip
  %% t3a_55_all.tex
$$
\ninepoint
\vbox{\hbox{\indent\vbox{
\tabskip=0pt\offinterlineskip
\def\tlr{\noalign{\hrule}}
\def\CO{{\cal O}}
\halign {\strut#& \vrule#\tabskip=2pt&
  \hfil$#$\hfil&\vrule\vrule#&
  \hfil$#$\hfil&\vrule#&
  \hfil$#$\hfil&\vrule#&
  \hfil$#$\hfil&\vrule#&
  \hfil$#$\hfil&\vrule\vrule#&
  \hfil$#$\hfil&\vrule#&
  \hfil$#$\hfil&\vrule#&
  \hfil$#$\hfil&\vrule#&
  \hfil$#$\hfil&\vrule#\tabskip=0pt\cr\tlr
\omit&height2pt&\omit&&\omit&&\omit&&\omit&&
\omit&&\omit&&\omit&&\omit&&\omit&\cr
&&
&&PS[0]SP
&&PW[0]WP
&&PS[0]WP
&&PS[1]WP
&&AS[0]SA
&&AW[0]WA
&&AS[0]WA
&&AS[1]WA
& \cr\tlr
\omit&height2pt&\omit&&\omit&&\omit&&\omit&&
\omit&&\omit&&\omit&&\omit&&\omit&\cr\tlr
&&
&&  2.6
&&  0.6
&&  0.5
&&  2.1
&&  1.9
&&  1.8
&&  0.9
&&  0.8
  &\cr
&&
&& 10-30
&& 16-27
&& 10-30
&& 23-33
&& 12-28
&& 14-28
&& 12-30
&& 26-33
  &\cr
&& {\cal P}^1
&&   -7.08( 97)
&&   -6.48( 41)
&&   -6.49( 39)
&&   -3.42( 57)
&&   -5.98( 71)
&&   -5.99( 44)
&&   -5.58( 42)
&&   -3.24( 42)
  &\cr\tlr
&&
&&  3.2
&&  4.4
&&  1.8
&&  1.4
&&  4.5
&&  3.4
&&  1.5
&&  2.3
  &\cr
&&
&& 10-30
&& 14-27
&& 11-29
&& 28-33
&& 10-30
&& 14-28
&& 12-30
&& 26-33
  &\cr
&& {\cal S}^1
&&   -1.01( 18)
&&   -0.92( 19)
&&   -0.94( 10)
&&   -0.48( 13)
&&   -0.85( 14)
&&   -0.89( 19)
&&   -0.85( 10)
&&   -0.48( 11)
  &\cr\tlr
&&
&&  5.9
&&  8.7
&&  3.1
&&  1.2
&&  4.6
&& 10.6
&&  3.3
&&  1.1
  &\cr
&&
&& 13-29
&& 13-27
&& 13-27
&& 25-33
&& 14-32
&& 10-30
&& 12-30
&& 27-34
  &\cr
&& {\cal V}_s^1
&&    0.83( 65)
&&    0.87( 41)
&&    0.91( 17)
&&    0.32( 18)
&&    0.83( 18)
&&    0.83( 31)
&&    0.79( 18)
&&    0.36( 27)
  &\cr\tlr
&&
&&  6.0
&&  2.9
&&  1.5
&&  1.1
&&  5.3
&&  1.5
&&  1.0
&&  0.5
  &\cr
&&
&& 15-26
&& 11-29
&& 14-26
&& 26-33
&& 11-30
&& 14-28
&& 13-29
&& 26-33
  &\cr
&& {\cal V}_t^1
&&    0.38( 15)
&&    0.45(  9)
&&    0.42(  5)
&&    0.22(  6)
&&    0.42(  9)
&&    0.41(  8)
&&    0.40(  5)
&&    0.23(  7)
  &\cr\tlr
&&
&&  4.4
&&  2.6
&&  1.6
&&  0.3
&&  0.7
&&  3.9
&&  2.2
&&  0.8
  &\cr
&&
&& 10-30
&& 12-28
&& 14-27
&& 27-33
&&  8-32
&& 12-28
&& 10-30
&& 26-33
  &\cr
&& {\cal A}_s^1
&&   -1.02( 17)
&&   -1.09( 20)
&&   -1.01( 12)
&&   -0.49( 15)
&&   -0.92( 12)
&&   -0.97( 19)
&&   -0.89( 10)
&&   -0.38( 14)
  &\cr\tlr
&&
&&  0.3
&&  4.3
&&  2.9
&&  3.0
&&  1.7
&&  4.2
&&  1.8
&&  2.1
  &\cr
&&
&& 12-28
&& 10-30
&& 12-28
&& 28-33
&& 10-28
&& 13-27
&& 14-27
&& 25-34
  &\cr
&& {\cal A}_t^1
&&    0.01(  5)
&&   -0.01(  3)
&&   -0.03(  3)
&&    0.12(  7)
&&   -0.01(  4)
&&    0.02(  5)
&&   -0.01(  3)
&&    0.14(  9)
  &\cr\tlr
&&
&&  3.9
&&  4.9
&&  0.9
&&  1.8
&&  1.2
&&  4.0
&&  1.1
&&  0.9
  &\cr
&&
&& 12-28
&& 12-30
&& 11-29
&& 26-33
&& 12-24
&& 14-28
&& 11-29
&& 26-33
  &\cr
&& {\cal T}_s^1
&&    5.66( 69)
&&    5.09( 59)
&&    5.27( 30)
&&    2.73( 52)
&&    4.90( 91)
&&    4.79( 66)
&&    4.75( 39)
&&    2.76( 47)
  &\cr\tlr
&&
&&  4.0
&&  2.9
&&  0.7
&&  2.3
&&  2.5
&&  3.1
&&  1.3
&&  1.5
  &\cr
&&
&& 12-28
&& 13-30
&& 12-28
&& 26-33
&& 13-25
&& 14-28
&& 10-30
&& 26-33
  &\cr
&& {\cal T}_t^1
&&    5.86( 77)
&&    5.39( 49)
&&    5.38( 32)
&&    3.24( 48)
&&    4.87(121)
&&    5.01( 46)
&&    4.93( 30)
&&    2.96( 48)
  &\cr\tlr
}}}}$$
  \par 
  {\bf Table 3a.} The same as  in Table  2a, but for  $\kappa = 0.155$. \vfill\endinsert

  \pageinsert\parindent=0pt %%%\baselineskip=1.5\normalbaselineskip
  %% t3b_55_all.tex

$$
\ninepoint
\vbox{\hbox{\indent\vbox{
\tabskip=0pt\offinterlineskip
\def\tlr{\noalign{\hrule}}
\def\CO{{\cal O}}
\halign {\strut#& \vrule#\tabskip=2pt&
  \hfil$#$\hfil&\vrule\vrule#&
  \hfil$#$\hfil&\vrule#&
  \hfil$#$\hfil&\vrule#&
  \hfil$#$\hfil&\vrule#&
  \hfil$#$\hfil&\vrule\vrule#&
  \hfil$#$\hfil&\vrule#&
  \hfil$#$\hfil&\vrule#&
  \hfil$#$\hfil&\vrule#&
  \hfil$#$\hfil&\vrule#\tabskip=0pt\cr\tlr
\omit&height2pt&\omit&&\omit&&\omit&&\omit&&
\omit&&\omit&&\omit&&\omit&&\omit&\cr
&&
&&PS[0]SP
&&PW[0]WP
&&PS[0]WP
&&PS[1]WP
&&AS[0]SA
&&AW[0]WA
&&AS[0]WA
&&AS[1]WA
& \cr\tlr
\omit&height2pt&\omit&&\omit&&\omit&&\omit&&
\omit&&\omit&&\omit&&\omit&&\omit&\cr\tlr
&&
&&  2.6
&&  2.7
&&  0.6
&&  1.0
&&  2.0
&&  1.6
&&  0.9
&&  0.8
  &\cr
&&
&& 10-30
&& 13-29
&& 10-30
&& 24-32
&& 12-28
&& 14-28
&& 11-29
&& 26-33
  &\cr
&& {\cal P}^2
&&  -19.7(26)
&&  -18.4(15)
&&  -18.0(11)
&&   -9.0(15)
&&  -16.7(23)
&&  -16.8(13)
&&  -15.8(12)
&&   -9.0(11)
  &\cr\tlr
&&
&&  2.7
&&  4.0
&&  2.4
&&  1.3
&&  2.2
&&  3.3
&&  1.1
&&  2.2
  &\cr
&&
&& 10-30
&& 13-27
&& 10-30
&& 26-35
&& 12-28
&& 14-28
&& 15-27
&& 26-33
  &\cr
&& {\cal S}^2
&&   -1.45( 32)
&&   -1.44( 33)
&&   -1.43( 17)
&&   -0.94( 23)
&&   -0.98( 31)
&&   -1.24( 14)
&&   -1.05( 24)
&&   -0.82( 15)
  &\cr\tlr
&&
&&  5.2
&&  3.9
&&  1.0
&&  1.4
&&  1.6
&&  2.5
&&  1.0
&&  1.6
  &\cr
&&
&& 11-29
&& 14-28
&& 12-26
&& 26-33
&& 14-27
&& 16-30
&& 12-28
&& 27-32
  &\cr
&& {\cal V}_s^2
&&    0.07(  4)
&&    0.06(  4)
&&    0.08(  2)
&&    0.00(  4)
&&    0.13( 11)
&&    0.02(  7)
&&    0.05(  4)
&&   -0.08(  9)
  &\cr\tlr
&&
&&  3.8
&&  3.4
&&  2.0
&&  1.4
&&  4.6
&&  4.3
&&  1.7
&&  1.3
  &\cr
&&
&& 12-28
&& 14-27
&& 10-30
&& 26-33
&& 14-27
&& 14-28
&& 15-27
&& 26-31
  &\cr
&& {\cal V}_t^2
&&    0.04(  1)
&&    0.04(  1)
&&    0.04(  1)
&&    0.01(  1)
&&    0.02(  3)
&&    0.04(  2)
&&    0.03(  1)
&&    0.03(  2)
  &\cr\tlr
&&
&&  7.4
&&  3.6
&&  2.5
&&  1.8
&&  2.8
&&  1.5
&&  1.4
&&  1.3
  &\cr
&&
&& 12-28
&& 11-31
&& 10-30
&& 26-35
&& 14-27
&& 15-27
&& 11-29
&& 26-34
  &\cr
&& {\cal A}_s^2
&&   -1.06( 17)
&&   -0.96(  6)
&&   -0.97(  9)
&&   -0.52(  9)
&&   -0.89( 14)
&&   -0.86(  8)
&&   -0.87( 13)
&&   -0.42(  8)
  &\cr\tlr
&&
&&  0.1
&&  3.9
&&  1.9
&&  1.2
&&  0.3
&&  5.9
&&  1.4
&&  1.1
  &\cr
&&
&& 12-28
&& 15-25
&& 16-24
&& 26-33
&& 12-28
&& 14-28
&& 12-25
&& 28-34
  &\cr
&& {\cal A}_t^2
&&    0.57( 10)
&&    0.51(  9)
&&    0.49(  9)
&&    0.61( 13)
&&    0.49( 10)
&&    0.48( 11)
&&    0.45(  4)
&&    0.53(  8)
  &\cr\tlr
&&
&&  4.0
&&  3.3
&&  2.6
&&  2.2
&&  4.2
&&  3.8
&&  4.1
&&  2.1
  &\cr
&&
&& 15-25
&& 13-27
&& 14-27
&& 28-34
&& 12-28
&& 14-28
&&  9-31
&& 26-33
  &\cr
&& {\cal T}_s^2
&&    0.14(  6)
&&    0.13(  3)
&&    0.14(  3)
&&    0.03(  2)
&&    0.09(  3)
&&    0.11(  3)
&&    0.11(  4)
&&    0.06(  3)
  &\cr\tlr
&&
&&  4.4
&&  1.6
&&  2.6
&&  1.0
&&  1.9
&&  3.0
&&  1.8
&&  0.8
  &\cr
&&
&& 14-27
&& 14-27
&& 12-28
&& 26-33
&& 14-26
&& 17-26
&& 10-30
&& 26-35
  &\cr
&& {\cal T}_t^2
&&    0.15(  4)
&&    0.14(  1)
&&    0.14(  2)
&&    0.05(  2)
&&    0.06(  5)
&&    0.12(  8)
&&    0.13(  4)
&&    0.07(  2)
  &\cr\tlr
}}}}$$
  \par 
  {\bf Table  3b.} The same as Table 3a, but for   the two  color loop
  contribution. \vfill\endinsert

  \pageinsert\parindent=0pt %%%\baselineskip=1.5\normalbaselineskip
  %% t4_gsq.tex
%\magnification=1200
$$
\vbox{\hbox{\indent\vbox{\tabskip=0pt\offinterlineskip
\def\tlr{\noalign{\hrule}}
\def\q{\quad}
\def\s{\phantom{-}}
\def\sq{\phantom{-}\q}
\def\Bpzero{{{B}^L_K(\vec p = 0)}}
\def\Bpone{{B}^L_K(\vec p = 1)}
\def\BK{{B}_K}

\halign {\strut#& \vrule#\tabskip=3pt&
  \hfil$#$\hfil&\vrule#&
  \hfil$#$\hfil&\vrule\vrule#&
  \hfil$#$\hfil&\vrule#&
  \hfil$#$\hfil&\vrule#&
  \hfil$#$\hfil&\vrule\vrule#&
  \hfil$#$\hfil&\vrule#&
  \hfil$#$\hfil&\vrule#&
  \hfil$#$\hfil&\vrule#\tabskip=0pt\cr\tlr
\omit&height4pt& \multispan{3} && \multispan{5} &&  \multispan{5} &\cr
&& \multispan{3} && \multispan{5} \hfil $\kappa=0.154$ \hfil
                 && \multispan{5} \hfil $\kappa=0.155$ \hfil &\cr
\omit&height4pt& \multispan{3} && \multispan{5} &&  \multispan{5} &\cr\tlr
\omit&height4pt&\omit&&\omit&&\omit&&\omit&&\omit&&\omit&&\omit&&\omit&\cr
&&g^2_{eff} && \mu a   &&  \Bpzero    && \Bpone      &&  \BK
                       && \Bpzero     &&  \Bpone     &&  \BK       &\cr
&&          &&         &&  12-28      &&  24-34      &&
                       &&  10-30      &&  26-34      &&            &\cr
\omit&height4pt&\omit&&\omit&&\omit&&\omit&&\omit&&\omit&&\omit&&\omit&\cr\tlr
\omit&height2pt&\omit&&\omit&&\omit&&\omit&&\omit&&\omit&&\omit&&\omit&\cr\tlr
\omit&height2pt&\omit&&\omit&&\omit&&\omit&&\omit&&\omit&&\omit&&\omit&\cr
&& 0.00   &&\s   -     &&\s 0.298(44) &&\s 0.446(42) &&  0.83(21)
                       &&\s-0.023(63) &&\s 0.263(62) &&  0.77(19)  &\cr
\omit&height2pt&\omit&&\omit&&\omit&&\omit&&\omit&&\omit&&\omit&&\omit&\cr\tlr
\omit&height2pt&\omit&&\omit&&\omit&&\omit&&\omit&&\omit&&\omit&&\omit&\cr
&& 1.00   &&\s 1.0     &&\s 0.373(45) &&\s 0.494(52) &&  0.81(23)
                       &&\s 0.109(74) &&\s 0.330(61) &&  0.71(20)  &\cr
\omit&height2pt&\omit&&\omit&&\omit&&\omit&&\omit&&\omit&&\omit&&\omit&\cr\tlr
\omit&height2pt&\omit&&\omit&&\omit&&\omit&&\omit&&\omit&&\omit&&\omit&\cr
&& 1.00   &&\s \pi     &&\s 0.362(42) &&\s 0.476(50) &&  0.78(22)
                       &&\s 0.110(71) &&\s 0.319(60) &&  0.69(22)  &\cr
\omit&height2pt&\omit&&\omit&&\omit&&\omit&&\omit&&\omit&&\omit&&\omit&\cr\tlr
\omit&height2pt&\omit&&\omit&&\omit&&\omit&&\omit&&\omit&&\omit&&\omit&\cr
&& 1.00   &&\s 1.7\pi  &&\s 0.356(42) &&\s 0.468(50) &&  0.76(22)
                       &&\s 0.110(70) &&\s 0.315(59) &&  0.67(21)  &\cr
\omit&height2pt&\omit&&\omit&&\omit&&\omit&&\omit&&\omit&&\omit&&\omit&\cr\tlr
\omit&height2pt&\omit&&\omit&&\omit&&\omit&&\omit&&\omit&&\omit&&\omit&\cr
&& 1.338  &&\s 1.0     &&\s 0.405(44) &&\s 0.512(55) &&  0.79(24)
                       &&\s 0.175(75) &&\s 0.359(69) &&  0.68(24)  &\cr
\omit&height2pt&\omit&&\omit&&\omit&&\omit&&\omit&&\omit&&\omit&&\omit&\cr\tlr
\omit&height2pt&\omit&&\omit&&\omit&&\omit&&\omit&&\omit&&\omit&&\omit&\cr
&& 1.338  &&\s \pi     &&\s 0.388(42) &&\s 0.486(53) &&  0.75(22)
                       &&\s 0.176(71) &&\s 0.344(62) &&  0.64(22)  &\cr
\omit&height2pt&\omit&&\omit&&\omit&&\omit&&\omit&&\omit&&\omit&&\omit&\cr\tlr
\omit&height2pt&\omit&&\omit&&\omit&&\omit&&\omit&&\omit&&\omit&&\omit&\cr
&& 1.338  &&\s 1.7\pi  &&\s 0.380(41) &&\s 0.475(51) &&  0.72(22)
                       &&\s 0.176(69) &&\s 0.337(61) &&  0.62(22)  &\cr
\omit&height2pt&\omit&&\omit&&\omit&&\omit&&\omit&&\omit&&\omit&&\omit&\cr\tlr
\omit&height2pt&\omit&&\omit&&\omit&&\omit&&\omit&&\omit&&\omit&&\omit&\cr
&& 1.75   &&\s 1.0     &&\s 0.453(45) &&\s 0.536(58) &&  0.75(23)
                       &&\s 0.268(70) &&\s 0.401(74) &&  0.64(25)  &\cr
\omit&height2pt&\omit&&\omit&&\omit&&\omit&&\omit&&\omit&&\omit&&\omit&\cr\tlr
\omit&height2pt&\omit&&\omit&&\omit&&\omit&&\omit&&\omit&&\omit&&\omit&\cr
&& 1.75   &&\s \pi     &&\s 0.429(42) &&\s 0.497(53) &&  0.68(22)
                       &&\s 0.268(63) &&\s 0.379(71) &&  0.57(23)  &\cr
\omit&height2pt&\omit&&\omit&&\omit&&\omit&&\omit&&\omit&&\omit&&\omit&\cr\tlr
\omit&height2pt&\omit&&\omit&&\omit&&\omit&&\omit&&\omit&&\omit&&\omit&\cr
&& 1.75   &&\s 1.7\pi  &&\s 0.417(41) &&\s 0.497(51) &&  0.64(21)
                       &&\s 0.268(59) &&\s 0.369(69) &&  0.55(22)  &\cr
\omit&height2pt&\omit&&\omit&&\omit&&\omit&&\omit&&\omit&&\omit&&\omit&\cr\tlr
}}}}$$
  \par 
  {\bf Table  4.}  The  lattice  $B-$parameter  for  the  perturbatively
  improved operator  $\hat{\cal O}$ for   zero and one  unit of  lattice
  momentum, and the $B$-parameter  after momentum subtraction.  The data
  show the magnitude  of the variation with the  value of $\gsqeff$ and
  $\mu a$ used in the  perturbative renormalization constants. The range
  of time-slices used in the fits is specified in the header. \vfill\endinsert

  \pageinsert\parindent=0pt %%%\baselineskip=1.5\normalbaselineskip
  %% t5_comp.tex
$$
\vbox{\hbox{\indent\vbox{\tabskip=0pt\offinterlineskip
\def\tlr{\noalign{\hrule}}
\def\q{\quad}
\def\s{\phantom{-}}
\def\sq{\phantom{-}\q}
\def\CVs{{\cal V}_s^1}
\def\CVS{{\cal V}_s^2}
\def\CVt{{\cal V}_t^1}
\def\CVT{{\cal V}_t^2}
\def\CAs{{\cal A}_s^1}
\def\CAS{{\cal A}_s^2}
\def\CAt{{\cal A}_t^1}
\def\CAT{{\cal A}_t^2}

\def\CA{B_{\cal A}}
\def\CV{B_{\cal V}}

\halign {\strut#& \vrule#\tabskip=3pt&
  \hfil$#$\hfil&\vrule\vrule#&
       $#$\hfil&\vrule#&
       $#$\hfil&\vrule\vrule#&
       $#$\hfil&\vrule#&
       $#$\hfil&\vrule#\tabskip=0pt\cr\tlr
\omit&height2pt&\omit&&\omit&&\omit&&\omit&&\omit&\cr
&&  && \hfil \hbox{Wilson} && \hfil \hbox{Staggered}
&& \hfil \hbox{Wilson} && \hfil \hbox{Staggered} &\cr
&&  && \kappa=0.154 &&  m_q=0.02+0.03  && \kappa=0.155 &&  m_q=0.01+0.02  &\cr
\omit&height2pt&\omit&&\omit&&\omit&&\omit&&\omit&\cr\tlr
\omit&height2pt&\omit&&\omit&&\omit&&\omit&&\omit&\cr\tlr
\omit&height2pt&\omit&&\omit&&\omit&&\omit&&\omit&\cr
&& M_K    &&\s 0.370(6)  &&\sq 0.374(3)  &&\s 0.297(11) &&\sq 0.296(2) &\cr
\omit&height2pt&\omit&&\omit&&\omit&&\omit&&\omit&\cr\tlr
\omit&height2pt&\omit&&\omit&&\omit&&\omit&&\omit&\cr
&& E_K    &&\s 0.511(12) &&\sq           &&\s 0.466(22) &&\sq          &\cr
\omit&height2pt&\omit&&\omit&&\omit&&\omit&&\omit&\cr\tlr
\omit&height2pt&\omit&&\omit&&\omit&&\omit&&\omit&\cr\tlr
\omit&height2pt&\omit&&\omit&&\omit&&\omit&&\omit&\cr
&& \CVs   &&  -0.57(33)  &&\q -0.48(2)   &&  -0.71(53)  &&\q -0.79(4)  &\cr
\omit&height2pt&\omit&&\omit&&\omit&&\omit&&\omit&\cr\tlr
\omit&height2pt&\omit&&\omit&&\omit&&\omit&&\omit&\cr
&& \CVS   &&  -0.11(8)   &&\q -0.036(4)  &&  -0.13(11)  &&\q -0.10(1)  &\cr
\omit&height2pt&\omit&&\omit&&\omit&&\omit&&\omit&\cr\tlr
\omit&height2pt&\omit&&\omit&&\omit&&\omit&&\omit&\cr
&& \CVt   &&  -0.10(7)   &&\q -0.056(3)  &&  -0.12(15)  &&\q -0.13(1)  &\cr
\omit&height2pt&\omit&&\omit&&\omit&&\omit&&\omit&\cr\tlr
\omit&height2pt&\omit&&\omit&&\omit&&\omit&&\omit&\cr
&& \CVT   &&\s 0.01(2)   &&\q -0.009(1)  &&  -0.02(4)   &&\q -0.024(2) &\cr
\omit&height2pt&\omit&&\omit&&\omit&&\omit&&\omit&\cr\tlr
\omit&height2pt&\omit&&\omit&&\omit&&\omit&&\omit&\cr
&& \CAs   &&\s 0.28(18)  &&\sq 0.15(1)   &&\s 0.41(42)  &&\sq 0.36(2)  &\cr
\omit&height2pt&\omit&&\omit&&\omit&&\omit&&\omit&\cr\tlr
\omit&height2pt&\omit&&\omit&&\omit&&\omit&&\omit&\cr
&& \CAS   &&\s 0.27(16)  &&\sq 0.063(5)  &&\s 0.29(32)  &&\sq 0.13(1)  &\cr
\omit&height2pt&\omit&&\omit&&\omit&&\omit&&\omit&\cr\tlr
\omit&height2pt&\omit&&\omit&&\omit&&\omit&&\omit&\cr
&& \CAt   &&\s 0.28(12)  &&\sq 0.33(1)   &&\s 0.40(20)  &&\sq 0.42(1)  &\cr
\omit&height2pt&\omit&&\omit&&\omit&&\omit&&\omit&\cr\tlr
\omit&height2pt&\omit&&\omit&&\omit&&\omit&&\omit&\cr
&& \CAT   &&\s 0.79(26)  &&\sq 0.81(1)   &&\s 0.82(40)  &&\sq 0.85(2)  &\cr
\omit&height2pt&\omit&&\omit&&\omit&&\omit&&\omit&\cr\tlr
\omit&height2pt&\omit&&\omit&&\omit&&\omit&&\omit&\cr\tlr
\omit&height2pt&\omit&&\omit&&\omit&&\omit&&\omit&\cr
&& \CA    &&\s 1.62(42)  &&\sq 1.35(3)   &&\s 1.92(89)  &&\sq 1.76(5)  &\cr
\omit&height2pt&\omit&&\omit&&\omit&&\omit&&\omit&\cr\tlr
\omit&height2pt&\omit&&\omit&&\omit&&\omit&&\omit&\cr
&& \CV    &&  -0.78(40)  &&\q -0.58(2)   &&  -0.98(71)  &&\q -1.04(5)  &\cr
\omit&height2pt&\omit&&\omit&&\omit&&\omit&&\omit&\cr\tlr
\omit&height2pt&\omit&&\omit&&\omit&&\omit&&\omit&\cr\tlr
\omit&height2pt&\omit&&\omit&&\omit&&\omit&&\omit&\cr
&& B_K    &&\s 0.83(21)  &&\sq 0.76(1)   &&\s 0.77(19)  &&\sq 0.72(2)  &\cr
\omit&height2pt&\omit&&\omit&&\omit&&\omit&&\omit&\cr\tlr
}}}}$$
  \par 
  {\bf  Table  5.}  Comparison  of    individual   $B-$parameters,   for
  space/time  and  1-loop/2-loop components   of the  vector   and axial
  four-fermion  operators,  between  Wilson  and  staggered  fermions at
  matching  values of  the kaon  mass.  The kaon   energy  at $\vec p  =
  (0,0,0)$ and $\vec p = (0,0,1)$ measured  from the 2-point correlators
  is also given. All results are quoted for $\gsqeff = 0$. \vfill\endinsert

  \pageinsert\parindent=0pt %%%\baselineskip=1.5\normalbaselineskip
  %% t6_gsq_D.tex
\magnification=1200
%% This data is after correcting for the factor of 4 on the
%% off-diagonal terms.

$$
\vbox{\hbox{\indent\vbox{\tabskip=0pt\offinterlineskip
\def\tlr{\noalign{\hrule}}
\def\phskip{\phantom{\kappa=0.154 \quad}}
\def\geff{{g^2_{eff}}}

\halign {\strut#& \vrule#\tabskip=3pt&
  \hfil$#$\hfil&\vrule#&
  \hfil$#$     &\vrule#&
  \hfil$#$     &\vrule#&
       $#$\hfil&\vrule#&
       $#$\hfil&\vrule#&
  \hfil$#$\hfil&\vrule#\tabskip=0pt\cr\tlr
\omit&height2pt& && && && && && &\cr
&& && B_K^L(0) && B_K^L(1) &&\hfil M_K &&\hfil E(1) && B_K    &\cr
\omit&height2pt& && && && && && &\cr\tlr
\omit&height2pt&\multispan{11}&\cr\tlr
%
% Beta = 5.5;   kappa = 0.159
% fit range 6 - 11 (program wf_fit)
%
\omit&height2pt& && && && && && &\cr
&&\phskip \geff=0.000          && 0.327(88)&& 0.507(96)&&
          &&          && 0.98(43)&\cr
&&\kappa=0.159\quad \geff=1.091&& 0.389(74)&& 0.537(80)&&
 0.477(13)&& 0.660(16)&& 0.92(36)&\cr
&&\phskip \geff=1.909          && 0.401(56)&& 0.505(58)&&
          &&          && 0.78(27)&\cr\tlr
%
% Beta = 5.5;   kappa = 0.160
% fit range 6 - 11 (program wf_fit)
%
&&\phskip           \geff=0.000&&-0.593(88)&&-0.052(51)&&
          &&          && 0.93(24)&\cr
&&\kappa=0.160\quad \geff=1.091&&-0.415(76)&& 0.051(44)&&
 0.362(7) && 0.562(22)&& 0.89(21)&\cr
&&\phskip           \geff=1.909&&-0.217(57)&& 0.132(34)&&
          &&          && 0.76(17)&\cr\tlr
%
% Beta = 5.6;   kappa = 0.156
% fit range 6 - 12 (program wf_fit)
%
\omit&height2pt&\multispan{11}&\cr\tlr
&&\phskip           \geff=0.000&& 0.497(33)&& 0.685(39)&&
          &&          && 1.23(20)&\cr
&&\kappa=0.156\quad \geff=1.071&& 0.534(31)&& 0.693(36)&&
 0.456(5) && 0.613(10)&& 1.15(19)&\cr
&&\phskip           \geff=1.875&& 0.512(26)&& 0.629(30)&&
          &&          && 0.97(16)&\cr\tlr
%
% Beta = 5.6;   kappa = 0.157
% fit range 6 - 11 (program wf_fit)
%
&&\phskip           \geff=0.000&& 0.154(25)&& 0.436(29)&&
          &&          && 1.14(19)&\cr
&&\kappa=0.157\quad \geff=1.071&& 0.161(25)&& 0.438(29)&&
 0.358(7) && 0.501(16)&& 1.13(19)&\cr
&&\phskip           \geff=1.875&& 0.234(21)&& 0.439(24)&&
          &&          && 0.95(16)&\cr\tlr
}}}}$$
  \par 
  {\bf  Table 6.} The same as in  Table 4, but for lattices  generated
  with  two flavors  of dynamical Wilson quarks.  The  upper and lower
  halves of  the   table  correspond to  $\beta=5.5$  and  $\beta=5.6$
  respectively.  These numbers are obtained with $\mu a = 1.0$. \vfill\endinsert

  \pageinsert\parindent=0pt %%%\baselineskip=1.5\normalbaselineskip
  %% t7_LR.tex
\magnification=1200
$$
\vbox{\hbox{\indent\vbox{\tabskip=0pt\offinterlineskip
\def\tlr{\noalign{\hrule}}
\def\q{\quad}
\def\s{\phantom{-}}
\def\sq{\phantom{-}\q}
\def\COseven{{\cal O}_7^{3/2}}
\def\COeight{{\cal O}_8^{3/2}}
\halign {\strut#& \vrule#\tabskip=3pt&
  \hfil$#$\hfil&\vrule#&
  \hfil$#$\hfil&\vrule\vrule#&
  \hfil$#$\hfil&\vrule#&
  \hfil$#$\hfil&\vrule\vrule#&
  \hfil$#$\hfil&\vrule#&
  \hfil$#$\hfil&\vrule#\tabskip=0pt\cr\tlr
\omit&height4pt& \multispan{3} && \multispan{3} && \multispan{3} &\cr
&& \multispan{3} && \multispan{3} \hfil $\kappa=0.154$ \hfil
                 && \multispan{3} \hfil $\kappa=0.155$ \hfil &\cr
\omit&height4pt& \multispan{3} && \multispan{3} && \multispan{3} &\cr\tlr
\omit&height4pt&\omit&&\omit&&\omit&&\omit&&\omit&&\omit&\cr
&&g^2_{eff}&& \mu a    &&  \COseven   && \COeight    && \COseven   &&
 \COeight   &\cr
\omit&height4pt&\omit&&\omit&&\omit&&\omit&&\omit&&\omit&\cr\tlr
\omit&height2pt&\omit&&\omit&&\omit&&\omit&&\omit&&\omit&\cr\tlr
\omit&height2pt&\omit&&\omit&&\omit&&\omit&&\omit&&\omit&\cr
&& 0.00   &&\s   -     &&\s 0.902(24) &&\s 0.947(30) &&\s 0.908(44)&&
\s 0.963(50)&\cr
\omit&height2pt&\omit&&\omit&&\omit&&\omit&&\omit&&\omit&\cr\tlr
\omit&height2pt&\omit&&\omit&&\omit&&\omit&&\omit&&\omit&\cr
&& 1.00   &&\s 1.0     &&\s 0.871(24) &&\s 0.918(29) &&\s 0.878(44)&&
\s 0.935(50)&\cr
\omit&height2pt&\omit&&\omit&&\omit&&\omit&&\omit&&\omit&\cr\tlr
\omit&height2pt&\omit&&\omit&&\omit&&\omit&&\omit&&\omit&\cr
&& 1.00   &&\s \pi     &&\s 0.918(29) &&\s 0.945(30) &&\s 0.911(46)&&
\s 0.962(52)&\cr
\omit&height2pt&\omit&&\omit&&\omit&&\omit&&\omit&&\omit&\cr\tlr
\omit&height2pt&\omit&&\omit&&\omit&&\omit&&\omit&&\omit&\cr
&& 1.00   &&\s 1.7\pi  &&\s 0.901(25) &&\s 0.953(30) &&\s 0.921(46)&&
\s 0.969(52)&\cr
\omit&height2pt&\omit&&\omit&&\omit&&\omit&&\omit&&\omit&\cr\tlr
\omit&height2pt&\omit&&\omit&&\omit&&\omit&&\omit&&\omit&\cr
&& 1.338  &&\s 1.0     &&\s 0.832(23) &&\s 0.877(23) &&\s 0.830(43)&&
\s 0.894(48)&\cr
\omit&height2pt&\omit&&\omit&&\omit&&\omit&&\omit&&\omit&\cr\tlr
\omit&height2pt&\omit&&\omit&&\omit&&\omit&&\omit&&\omit&\cr
&& 1.338  &&\s \pi     &&\s 0.891(25) &&\s 0.934(30) &&\s 0.902(46)&&
\s 0.951(51)&\cr
\omit&height2pt&\omit&&\omit&&\omit&&\omit&&\omit&&\omit&\cr\tlr
\omit&height2pt&\omit&&\omit&&\omit&&\omit&&\omit&&\omit&\cr
&& 1.338  &&\s 1.7\pi  &&\s 0.908(25) &&\s 0.950(30) &&\s 0.921(47)&&
\s 0.966(52)&\cr
\omit&height2pt&\omit&&\omit&&\omit&&\omit&&\omit&&\omit&\cr\tlr
\omit&height2pt&\omit&&\omit&&\omit&&\omit&&\omit&&\omit&\cr
&& 1.75   &&\s 1.0     &&\s 0.747(21) &&\s 0.785(26) &&\s 0.753(40)&&
\s 0.802(45)&\cr
\omit&height2pt&\omit&&\omit&&\omit&&\omit&&\omit&&\omit&\cr\tlr
\omit&height2pt&\omit&&\omit&&\omit&&\omit&&\omit&&\omit&\cr
&& 1.75   &&\s \pi     &&\s 0.869(25) &&\s 0.911(30) &&\s 0.883(46)&&
\s 0.928(51)&\cr
\omit&height2pt&\omit&&\omit&&\omit&&\omit&&\omit&&\omit&\cr\tlr
\omit&height2pt&\omit&&\omit&&\omit&&\omit&&\omit&&\omit&\cr
&& 1.75   &&\s 1.7\pi  &&\s 0.901(26) &&\s 0.941(31) &&\s 0.916(48)&&
\s 0.958(52)&\cr
\omit&height2pt&\omit&&\omit&&\omit&&\omit&&\omit&&\omit&\cr\tlr
}}}}$$
  \par 
  {\bf  Table 7.} The $B$-parameter for the LR electromagnetic penguin
  operators on the quenched lattices.  The fit range is $t=13-27$ in all
  cases. The results are shown for a number of values of $g^2_{eff}$
  and $\mu a$ used in the perturbative renormalization constants.  \vfill\endinsert

% References %%%%%%%%%%%%%%%%%%%%%%%%%%%%%%%%%%%%%%%%%%%%%%%%%%%%%%%%%
\listrefs
\vfill

\end